\documentclass[letter,12pt]{article}
\pdfoutput=1 

\usepackage{jheppub} 

\usepackage[utf8]{inputenc}

\usepackage[T1,T2A]{fontenc} 

\usepackage{amsmath}
\DeclareMathOperator{\Tr}{Tr}
\usepackage{ upgreek }

\def\bea{\begin{eqnarray}}
\def\eea{\end{eqnarray}}

\title{Spectrum of quantum KdV hierarchy in the semiclassical limit}

\author[a]{ Anatoly Dymarsky,} 
\author[a]{Ashish Kakkar,}
\author[]{Kirill Pavlenko,}
\author[a, b, c]{ and Sotaro Sugishita}

\affiliation[a]{Department of Physics and Astronomy, \\ University of Kentucky,\\Lexington, KY, USA 40506\\}
\affiliation[b]{Department of Physics, Nagoya University, Nagoya, Aichi 464-8602, Japan\\}
\affiliation[c]{Institute for Advanced Research, Nagoya University, Nagoya, Aichi 464-8601, Japan}

\abstract{
We employ semiclassical quantization to calculate spectrum of quantum KdV charges in the limit of large central charge $c$. Classically, KdV charges  $Q_{2n-1}$ generate completely integrable dynamics on the co-adjoint orbit of the Virasoro algebra. They can be  expressed in terms of action variables $I_k$, e.g.~as a power series expansion. Quantum-mechanically this series becomes the expansion in $1/c$, while action variables become integer-valued quantum numbers $n_i$. Crucially, classical expression, which is homogeneous in $I_k$, acquires quantum corrections that include terms of subleading  powers in $n_k$. At first two non-trivial orders in $1/c$ expansion   these ``quantum'' terms can be  fixed from the analytic form of $Q_{2n-1}$ acting on the primary states.  In this way we  find explicit expression for the spectrum of $Q_{2n-1}$ up to first  three orders in $1/c$ expansion. We apply this result to study thermal expectation values of $Q_{2n-1}$ and free energy of the KdV Generalized Gibbs Ensemble. 
}

\begin{document} 
\maketitle
\flushbottom

\section{Introduction}
\label{sec:intro}
Conformal invariance in two dimensions is a very powerful tool which gives rise to many non-pertubative relations constraining dynamics of 2d CFTs. Among them is universality of stress-energy tensor sector \cite{BPZ}, namely any correlation function which includes only stress-energy tensor and its descendants depends only on central charge $c$ but not on any other details of the theory. An analytic form of all such correlators can in principle be found  in a recursive form \cite{zamolodchikov1989conformal}.
The stress-energy sector can be regarded as integrable, even if the whole theory is understood to be chaotic \cite{hartman2014universal}.
This can be justified formally by noting there is an infinite number of mutually commuting quantum KdV charges \cite{bazhanov1996integrable, bazhanov1997integrable, bazhanov1999integrable} -- local charges $Q_{2n-1}$ of the form
\bea
Q_{2n-1}={1\over 2\pi}\int_0^{2\pi} T_{2n}(\varphi)\, d\varphi \label{qKdV},
\eea
where the densities $T_{2n}$ are appropriately regularized polynomials in stress-energy tensor $T(\varphi)$ and its derivatives. First charge 
\bea
Q_1=L_0-{c\over 24}={1\over 2\pi}\int_0^{2\pi} T\, d\varphi \label{Q1definition}
\eea
is the CFT Hamiltonian. (Here and below we consider 2d CFT on a cylinder. Because of  standard factorization  into left and right-moving sectors we restrict the discussion to one sector only.) Interest in integrable structure of 2d CFT stress-energy sector has been reignited recently in the context of Eigenstate Thermalization Hypothesis (ETH) \cite{srednicki1994chaos}. Following original works \cite{lashkari2018eigenstate, lin2016thermality,basu2017thermality,He2017txy,He2017vyf,lashkari2018universality,Guo2018pvi,maloney2018generalized,GGE} it has been conjectured  and  confirmed in \cite{GETH} that 2d CFTs exhibit generalized ETH with the local equilibrium being described by qKdV Generalized Gibbs Ensemble (GGE). Schematically the role of qKdV charges is as follows. The CFT Hamiltonian \eqref{Q1definition} is highly degenerate 
with all CFT descendant states of the form 
\bea
\label{E}
|E\rangle=L_{-m_1} \dots L_{-m_k} |\Delta \rangle,\qquad \sum_{i=1}^k m_i=m
\eea
sharing the same energy $E=\Delta+m-c/24$. Since all $Q_{2n-1}$ commute, they can be simultaneously diagonalized giving rise to mathematically unique ``integrable'' basis of eigenstates. Unlike the energy eigenstates of the form \eqref{E}, which fail the ETH, integrable eigenstates carry specific values of $Q_{2k-1}$-charges and obey generalized ETH. 
This novel role of qKdV symmetries  motivates  the question of ``solving'' integrable structure, i.e.~evaluating spectrum of qKdV charges and finding integrable eigenstates, which would allow detailed studies of generalized ETH and qKdV GGE thermodynamics. 

In certain sense the question of finding  qKdV spectra can be regarded as solved: there is not one but two distinct ways to write an algebraic Bethe-ansatz reducing the problem of finding spectra to a bunch of algebraic equations \cite{BazhanovEq, LitvinovEq}. In practice complexity of these equations grows very rapidly with the level $m$ \eqref{E}, making this approach  useless in the context of ETH, at least so far.  The ETH holds in thermodynamic limit, it may not and does not hold beyond that regime. Thermodynamic limit assumes the length of the spatial circle $L$ goes to infinity, with the energy density $E/L$ kept fixed. Using rescaling, one can always bring the circle to unit radius, the notations we use throughout the paper. The energy $E$ then must go to infinity as $L^2$ with $L\rightarrow \infty$ being an auxiliary parameter keeping track of corrections to various ETH-related identities. For any given primary state $|\Delta\rangle$ this essentially means the descendant level  $m$ must be taken to infinity, i.e.~we arrive exactly at the limit where algebraic Bethe equations become most difficult. 

A   progress was achieved by taking an additional limit of large central charge. In this case $Q_{2k-1}$-eigenstates, akin to \eqref{E},  can be parametrized by a set of natural numbers, which can be conveniently combined into a  Young tableau \cite{GGE}.\footnote{Appearance of $n_k$ to parametrize the eigenstates
 can be understood from the  Virasoro algebra, which in the large $c$ limit reduces to  a product of Heisenberg algebras, with $n_k$ being the corresponding quantum numbers  \cite{Witten,Brehm_2020}.} It is most convenient to use representation when $n_k \geq 0$ for $k=1,2,\dots$ counts the number of rows of length $k$, 
\bea
|n_i\rangle\equiv |n_1,\dots\rangle, \qquad \sum_{k=1}^\infty k\, n_k =m. \label{Qeig}
\eea 
We emphasize \eqref{Qeig} are eigenstates of $Q_{2n-1}$ and thus differ from \eqref{E}. Corresponding eigenvalues at leading order were conjectured in \cite{GGE2}\footnote{Since Ref.~\cite{GGE2} was working in the regime of both large central charge and thermodynamic limit $Q_1 \propto c L^2$, it only  conjectured the term linear in $n_k$, as the $n_k$-independent term is $1/L^2$ suppressed.}
\bea
Q_{2n-1}|n_i\rangle&=&{\sf Q}_{2n-1}|n_i\rangle,\\
\label{leadingQ}
{\sf Q}_{2n-1}&=&\tilde{\Delta}^n+\sum_{p=0}^{n-1} \xi_n^p\, \tilde{\Delta}^{n-1-p}\, \tilde{c}^p  \left( \sum_{k=1}^\infty k^{2p+1} n_k+{\zeta(-2p-1)\over 2}\right)+O(\tilde{c}^{n-2}), \nonumber \\
\xi_n^p&=&{(2n-1)\sqrt{\pi}\,\Gamma(n+1)\over \, 2\, \Gamma(p+3/2)\Gamma(n-p)},\qquad \tilde{\Delta}=\Delta-{\tilde c},\qquad \tilde{c}={c-1 \over 24}. \label{xidef}
\eea
Here we  assume the scaling when $c\rightarrow \infty$ while $\tilde{\Delta}/\tilde{c}=h$ is kept fixed. No thermodynamic limit is assumed. This is the limit  of holographic correspondence, when CFT is dual to semiclassical gravity. The holographic picture provides an easy derivation for the leading $1/c$ terms in  \eqref{leadingQ} and provides interpretation for $n_k$ as the boson occupation numbers of the boundary gravitons, see Appendix \ref{sec:holography}. 
From the mathematical point of view simplicity of eigenstates parametrization with help of Young tableaux as well as relatively simple form of \eqref{leadingQ} can be readily understood from the semiclassical quantization of the co-adjoint orbit of Virasoro algebra. Indeed, as is explained in \cite{Witten}  in the large $c$ limit Virasoro algebra can be understood in quasi-classical terms, as quantization of the Kirillov-Kostant-Souriau symplectic form. Because of $U(1)$ symmetry semiclassical quantization of $Q_1$ is exact, 
\bea
\label{Q1}
Q_1=\tilde{\Delta} +\left(\sum_{k=1}^\infty k\, n_k-{1\over 24}\right) = \Delta+m-{c\over 24},
\eea
but for all higher $Q_{2n-1}, n>1$ it is not. It is a perturbation series in $1/\tilde{c}$, which plays the effective role of Planck constant. In this paper we develop a perturbative scheme to obtain the spectrum of $Q_{2n-1}$ as a series in $1/\tilde{c}$ expansion and calculate first two non-trivial terms. The result is summarized  in \eqref{result}.

 In the strict $c\rightarrow \infty$ limit when the problem becomes classical, CFT stress-tensor $T$ can be substituted by an element of the co-adjoint orbit of Virasoro algebra ${24\over c}u$, where $u$ is a potential of an auxiliary periodic Schr$\ddot{\rm o}$dinger equation. Then quantum KdV charges \eqref{qKdV} reduce to conventional KdV Hamiltonians  of the periodic problem
\bea
Q_{2n-1}={1\over 2\pi}\int_0^{2\pi} (u^n+\dots )\, d\varphi,
\eea
which we denote the same as the quantum ones, as it clear from the constant which, classical or quantum version, we had in mind. 
For the states with large but finite level $m$ number of non-zero $n_k$ will also be finite. At the classical level this corresponds to finite-zone potentials $u$, which form a finite-dimensional symplectic  manifold equipped with the structure of a completely integrable system.   Hamiltonians ${Q}_{2n-1}$ can be re-expressed in terms of the action variables $I_k$ and the orbit invariant $h$, 
\bea
\label{tQ}
Q_{2n-1}=h^n+  \sum_{k=1}^\infty \sum_{j=0}^{n-1} \xi_n^j\, h^{n-1-j}  k^{2j+1}\, I_k+O(I^2) \label{qs}
\eea
which at semiclassical level become integral quantum numbers $I_k \rightarrow {n_k/\tilde{c}}$.
It is then easy to see that \eqref{qs} becomes \eqref{leadingQ}, up to an overall factor $\tilde{c}^n$ and certain corrections. At each power of $1/\tilde{c}$ classical expression $Q_{2n-1}(h,I_k)$ predicts only leading power of $n_k$ while all subleading powers are ``quantum corrections'' which must be fixed separately. 

At leading $1/\tilde{c}$ order  quantum correction is just $n_k$-independent constant term proportional to $\zeta(-2p-1)/2$, see \eqref{leadingQ}. It can be fixed trivially by introducing Maslov index $I_k \rightarrow (n_k+1/2)/\tilde{c}$, such that constant term can be formally rewritten as the vacuum energy of ``quantum oscillators'' with frequencies $\omega_k$ and 
 occupation numbers $n_k$
\bea
{\sf Q}_{2n-1}={\tilde \Delta}^n+ \sum_{k=1}^\infty (n_k+1/2)\, \omega_k+O(\tilde{c}^{n-2}),\quad \omega_k=\sum_{j=0}^{n-1}\xi_n^j\, \tilde{\Delta}^{n-1-j}\, \tilde{c}^j\, k^{2p+1}.
\eea
Unfortunately this simple trick fails beyond the leading order in $1/\tilde{c}$.  At $1/\tilde{c}^2$ order one  has to fix both constant and linear in $n_k$ terms, while   simple $I_k \rightarrow (n_k+1/2)/\tilde{c}$ substitution leads to incorrect results. 

We propose and verify up to  $1/\tilde{c}^2$ order that the subleading ``quantum correction'' terms can be unambiguously fixed starting from the analytic expression in terms of $1/\tilde c$ pertubative series of the eigenvalues ${\sf Q}^0_{2n-1}$   of $Q_{2n-1}$ acting on the primary state $|\Delta\rangle$. For leading $1/\tilde c$ term this statement is trivial -- taking all $n_k=0$ yields the constant term, which is simply leading $1/\tilde{c}$ term in ${\sf Q}^0_{2n-1}$. At the $1/\tilde  c^2$ order this statement is more nuanced:
naively ${\sf Q}^0_{2n-1}$  only fixes the  constant term with all $n_k=0$, but we show linear in $n_k$ terms can be also fixed starting from  ${\sf Q}^0_{2n-1}$.  As a result we obtain spectrum of $Q_{2n-1}$ up to first three orders in $1/ c$ expansion, including the leading $\Delta^n$ term. We then apply the obtained result to evaluate thermal expectation values of $Q_{2n-1}$, free energy of the KdV Generalized Gibbs Ensemble, and the asymptotic expansion of the quantum transfer matrix acting on a primary state $|\Delta\rangle$, all at first few leading orders in $1/c$. 

The paper is organized as follows. In section \ref{sec:classic} we discuss classic completely integrable system associated with the finite zone potentials and evaluate $Q_{2n-1}(h,I_k)$ as a perturbative series in $I_k$. In section \ref{sec:vacuumenergy} we discuss analytic form of $Q_{2k-1}$ acting on primary states. These two pieces are combined in section \ref{sec:quantization} where we employ semiclassical quantization  to obtain the spectrum of qKdV charges in  the first three orders of  $1/\tilde c$ expansion.
We also perform consistency checks, confirming our result. Section \ref{sec:applications} is devoted to applications of the obtained result. In section \ref{sec:thermalev} we calculate thermal expectation values of $Q_{2n-1}$ and fix two leading orders in $1/c$ of  the associated  differential operator ${\mathcal D}_n$
\bea
{\rm Tr}_\Delta (Q_{2k-1}\, q^{Q_1})={\mathcal D}_n \chi_\Delta,\qquad \chi_\Delta\equiv {\rm Tr}_\Delta (q^{Q_1}).
\eea 
In section \ref{sec:GGE} we discussed KdV Generalized Gibbs Ensemble and calculate its free energy $-\ln Z_{\rm GGE}$,
\bea
Z_{\rm GGE}={\rm Tr}\, e^{-\sum_n \mu_{2n-1}\, Q_{2n-1}},
\eea
at leading order in $1/c$. In section \ref{sec:TM} we use the asymptotic expansion to calculate the quantum transfer matrix acting on a primary state at first two orders in $1/c$ expansion. We use analytic continuation to extend the validity beyond the asymptotic regime, but notice that certain non-pertubative terms are missing. 
We conclude with a discussion in section \ref{sec: discussion}. The paper also includes a number of appendices. Appendix \ref{sec:holography} provides an easy derivation of \eqref{leadingQ} by quantizing boundary gravitons of semiclassical gravity in AdS${}_3$. Appendix \ref{sec:pert} evaluates $Q_{2n-1}(h,I_k)$ at first two orders in $I_k$ by  explicitly  introducing normal coordinates at the origin of the co-adjoint orbit of the Virasoro algebra. Appendix \ref{A:1cut} provides technical details concerning  Novikov's one-zone potentials. Appendix \ref{perturbative} develops the technique of dealing with the multi-zone potentials in the limit of the infinitesimally small zones. Finally, appendix \ref{qvac} provides the details of calculating the spectrum of $Q_{2n-1}$ acting on primary states based on ODE/IM correspondence.

\section{Calculation of $Q_{2n-1}(h,I_k)$}
\label{sec:classic}
In this section our goal is to find expression for $Q_{2n-1}$ in terms of the orbit invariant $h$ and action variables $I_k$, by expanding pertubatively up to cubic order in $I_k$, 
\begin{align}
\label{genQ}
    Q_{2n-1}&=h^n + \sum_k f^{(n,1)}_k I_k+f^{(n,2)}_{k}I_k^2+f^{(n,3)}_{k}I_k^3
+\sum_{\substack{k < \ell}} f^{(n,2)}_{k,\ell} I_k I_\ell + \\
 &\qquad\quad  \, \,\, \sum_{k\neq \ell} f^{(n,3)}_{k,\ell}I_k^2 I_\ell +\sum_{k<\ell<p} f^{(n,3)}_{k,\ell,p} I_k I_\ell I_p
    +\mathcal{O}(I^4).\nonumber
\end{align}
Coefficients $f$ are $h$-dependent. First three $f^{(n,1)}_k,f^{(n,2)}_k,f^{(n,3)}_k$ will be found using one-zone potentials in section \ref{1cut}. Using two-zone potentials we will find $f^{(n,2)}_{k,\ell}$ and $ f^{(n,3)}_{k,\ell}$ in section \ref{2cut}, while coefficient $f^{(n,3)}_{k,\ell,p}$ will be fixed using three-zone potentials in section \ref{3cut}.
An alternative brute-force derivation of \eqref{genQ} up to quadratic order in $I_k$  is given in the appendix \ref{sec:pert}.

\subsection{Finite zone potentials: an introduction}
The starting point is the ``Schr$\ddot{\rm o}$dinger'' equation 
\bea
-\psi''+{u\over 4}\, \psi=\lambda\, \psi, \label{sch}
\eea
with the periodic real-valued potential $u(\varphi+2\pi)=u(\varphi)$.
For any real $\lambda$ there are two linearly-independent quasi-periodic  solutions 
\bea
\psi_\pm (\varphi+2\pi)= e^{\pm 2\pi i\, p(\lambda)}\psi_\pm (\varphi).
\eea
Here quasi-momentum $p(\lambda)$ could be either real or pure imaginary. 
Values of $\lambda\in {\mathbb R}$ for which $p(\lambda)$ is imaginary are called ``forbidden zone.'' At the end of forbidden zones $p(\lambda)$ is integer or half-integer such that $\psi_\pm$ become  periodic or antiperiodic and linearly dependent. Normally, for such $\lambda$, another linearly independent singular solution  appears. 
Yet occasionally there are two linearly independent  regular periodic or antiperiodic solutions for the same $\lambda$. In this case forbidden zone degenerates and disappears, with $p(\lambda)$ being real everyone in the vicinity of that point. We provide examples below. 

A general potential $u$ would have an  infinite number of forbidden zones, but there are special classes when only a finite number of forbidden zones are non-degenerate, Such $u$ are called finite zone potentials. They were introduced in a famous work \cite{novikov1974periodic} and often refereed to  as Novikov potentials. 

\subsection*{Example: zero zone potential}
Let us consider a constant potential $u=4\lambda_0=Q_1$ with some real $Q_1$. A solution to \eqref{sch} can be readily found 
\bea
\label{s1}
\psi_\pm(\varphi)=e^{\pm ip(\lambda)\varphi},\qquad p(\lambda)=\sqrt{\lambda-\lambda_0}.
\eea
For any $\lambda>Q_1/4$ quasi-potential is real, i.e.~there are no forbidden zones, except for $\lambda \in (-\infty; Q_1/4)$. The solutions \eqref{s1} are linearly independent, including  $\lambda=(Q_1+k^2)/4$ for natural $k$, when $\psi_\pm $ are (anti)periodic. Values $\lambda=(Q_1+k^2)/4$ mark the ends of degenerate forbidden zones. 

\subsection*{Example: ``opening'' a zone}
Let us now consider the potential $u=Q_1+\epsilon \cos(k \varphi)+O(\epsilon^2)$ where $Q_1$ is a constant, $k$ is positive integer, and $\epsilon$ is some infinitesimal parameter. Using quantum mechanics perturbation theory we find at leading order that all eigenvalues of periodic and anti-periodic problems remain the same and double-degenerate, except for $\lambda_k$ which splits into 
\bea
\lambda_{k}^\pm = {Q_1+k^2\over 4} \pm {\epsilon\over 2}.
\eea
Hence now there are two forbidden zones, $(-\infty, Q_1/4)$ and $(\lambda_k^-,\lambda_k^+)$.\\

Finite-zone potentials are characterized by the ends of non-degenerate zones $\lambda_i$. For the zero-zone potential above there is only one parameter $\lambda_0=Q_1/4$. After one zone is opened, there are three parameters: ``energy'' of the ground state  $\lambda_0$, $\lambda_1=\lambda_{k}^-$ and $\lambda_2=\lambda_{k}^+$.
In general an $m$-zone potential is characterized by 
\bea
\lambda_0< \lambda_1 < \dots <\lambda_{2m},
\eea
with the forbidden zones $(-\infty, \lambda_0)$ and $(\lambda_{2i-1},\lambda_{2i})$, $i=\overline{1,m}$. For each set $\{\lambda_i\}$ we can define a hyperelliptic curve 
\bea
\label{curve}
y^2=\prod_{i=0}^{2m}(\lambda-\lambda_i),
\eea
while the quasi-momentum $p$ being fixed in terms of its differential
\bea
\label{dp}
dp={\lambda^m+r_{m-1}\lambda^{n-1}+\dots r_{0}\over 2\, y}d\lambda,\qquad p(\lambda_0)=0.
\eea
The latter is defined in such a way that the integrals of $dp$ over $a$-cycles vanish 
\bea
\label{zero}
\oint\limits_{a_i} dp=2\int_{\lambda_{2i-1}}^{\lambda_{2i}} dp =0.  \label{C1}
\eea
This fixes $m$ coefficients $r_0, \dots, r_{m-1}$. Furthermore for the potential associated with $\{\lambda_i\}$ to be $2\pi$-periodic we must additionally require integrals over $b$-cycles
\bea
\label{integerc}
w_i=\oint_{b_i}dp=2\int_{\lambda_{2i-2}}^{\lambda_{2i-1}} dp
\eea
to be integer-valued  
\bea
\label{periodicity}
w_i=k_i-k_{i-1}. \label{C2}
\eea 
Here natural $k_i$ satisfying  $k_{i+1}>k_i$, $k_0\equiv 0$, label opened zones. 
These are additional $m$ constrains, which reduce the total number of independent parameters $\lambda_i$ to $m+1$. 

A given set $\{\lambda_i\}$ which satisfies (\ref{C1},\ref{C2}), such that only $m+1$ parameters are independent, defines periodic potential $u(\varphi)$, but in a non-unique way. Individual potentials are labeled by points of the Jacobian of curve \eqref{curve}, with all of them sharing the same spectrum. In other words isospectral potentials form an $m$-dimensional torus, while  full space of $m$-zone potentials is therefore $2m+1$ dimensional. 

At this point we would like to make a connection with the Virasoro algebra. Consider Hill's equation, which is ``Schrodinger'' equation \eqref{sch} with $\lambda=0$, 
\bea
-\psi''+{u\over 4}\, \psi=0.
\eea
One can re-parametrize the circle going from $\varphi$ to $\tilde\varphi(\varphi)$ such that $\tilde\varphi(\varphi+2\pi)=\tilde\varphi(\varphi)+2\pi$. Then wave-function and the potential also change accordingly 
\bea
\label{psirep}
\tilde{\psi}(\tilde{\varphi})&=&\psi(\varphi) \left({d\tilde{\varphi}\over d\varphi}\right)^{-1/2},
\\
\tilde{u}(\tilde{\varphi})&=&\left( {d\tilde{\varphi}\over d\varphi} \right)^{-2} 
\left(u+2(S\tilde{\varphi})(\varphi)\right), \label{change}
\eea
where Schwarzian derivative 
\bea
(S\theta)(\varphi)\equiv {\theta'''\over \theta'}-{3\over 2}\left({\theta''\over \theta'}\right)^2.
\eea
From \eqref{change} it is clear that $u$ is an element from the co-adjoint orbit of Virasoro algebra with the Schwarzian derivative term appearing because of  central extension \cite{Witten}. All potentials $u(\varphi)$ related by circle reparametrizations, i.e.~belonging to the same co-adjoint orbit share the same invariant -- quasi-momentum at zero, 
\bea
\psi(2\pi)/\psi(0)=e^{2\pi ip(0)},
\eea
which is evident from \eqref{psirep}.
In other words 
\bea
\label{hdef}
-4p(0)^2=h
\eea
is the invariant of $u$ characterizing the orbit itself. By choosing an appropriate $\tilde \varphi$  the potential always\footnote{An implicit assumption here is that $u$ belongs to the regular orbit ${\rm diff}(\mathbb{S}^1)/\mathbb{S}^1$, which upon quantization, becomes Verma module. } can be brought to a constant form, in which case
\bea
\tilde{u}(\tilde{\varphi})=h.
\eea

The co-adjoint orbit is a symplectic space equipped with the Kirillov-Kostant-Souriau bracket 
\bea
\label{bracket}
{c\over 24}\{u(\varphi),u(\varphi')\}=-2\pi {\mathcal D}\delta(\varphi-\varphi'),\qquad 
{\mathcal D}=\partial u+2u \partial-2\partial^3.
\eea
Here, using linearity of symplectic form we introduce a formal parameter $c$, which later will be identified with the CFT central charge. 
Any Hamiltonian flow defined by \eqref{bracket} leaves $h$ invariant.

There is an infinite tower of the so-called KdV Hamiltonians $Q_{2k-1}$, which can be defined recursively  with help fo Gelfand-Dikii polynomials $R_n$,
\bea
\label{iterativeR}
&&Q_{2n-1}={1\over 2\pi}\int_0^{2\pi} R_{n}d\varphi\,\qquad \qquad \partial R_{n+1}={n+1\over 2n+1}{\mathcal D}R_{n},\quad  \\
&&R_0=1,\quad R_1=u,\quad R_2=u^2-{4\over 3}\partial^2 u,\quad R_3=u^3-4u \partial^2 u -2(\partial u)^2+{8\over 5}\partial^4 u,\dots \nonumber
\eea
Their Hamiltonian flows generate isospectral deformations of $u$ 
\bea
\delta u={c\over 24}\{Q_{2n-1},u\}=(2n-1)\partial R_n,
\eea
while they all remain in involution $\{Q_{2n-1},Q_{2\ell-1}\}=0$.

We now consider a space of all $m$-zone potentials sharing the same $h$. This is a $2m$-dimensional subspace within the orbit parametrized by $h$, which we will denote as ${\mathcal F}_m(h)$. The pullback of the symplectic form on this space is non-degenerate, hence it is also a  symplectic manifold  equipped with the Poisson bracket. Isospectral flows leave this manifold invariant.
Upon restricting to ${\mathcal F}_m(h)$, only first $n$  KdV Hamiltonians remain algebraically independent. The flows they generate move $u$ along the Jacobian of \eqref{curve}, which is the Liouvillian torus of a completely integrable dynamical system defined by $Q_{2n-1}$, $n\leq m$. In other words the geometry of ${\mathcal F}_m(h)$ is a $m$-dimensional torus parametrized by angle variables fibered above a base parametrized by $m$ variables $Q_{2n-1}$. Alternatively, one can introduce $m$ action variables $I_k$ parameterizing the base and forming canonical conjugate pairs with angle variables. 

In terms of $dp$ \eqref{dp} values of KdV charges are given by an expansion at infinity 
\bea
\label{kdveq}
Q_{2n-1}={2 \Gamma(n+1)\Gamma(1/2)\over \Gamma(n+1/2)} {4^n\over 2\pi i} \oint\limits_\infty dp\, \lambda^{n-1/2},
\eea
while the action variables are 
\bea
\label{action}
I_k={i\over \pi}\oint\limits_{a_k} p {d\lambda\over \lambda} ={1\over i \pi}\oint\limits_{a_k} dp \ln \lambda.
\eea
Functional dependence of $Q_{2n-1}$ for $n> m$ on the first $m$ ones readily follows from \eqref{kdveq} and the form of $dp$ \eqref{dp}.

Our task is conceptually trivial: we want to learn an explicit change of variables on the base of ${\mathcal F}_m(h)$ from $Q_{2n-1}$ to $I_k$. The expressions for $Q_{2n-1}(h,I_k)$ is not available in the closed form, we therefore will find  first few orders by expanding it in powers of $I_k$. 
There is one notable exception, using Riemann bilinear relation  with two one-forms $dp$ and $p d\lambda/\lambda$ one can show in full generality
\bea
\label{q1}
Q_1=h+\sum_k k\, I_k. 
\eea

Our main approach will be based on parameterizing both $Q_{2n-1}$ and $I_k$ in terms of the spectral curve $i=\overline{0,m}$, with the infinitesimal $\lambda_{2i}-\lambda_{2i-1}$, and then re-expressing $Q_{2n-1}$ in terms of $I_k$. There is an alternative straightforward approach, to parametrize the potential $u(\varphi)$ in terms of its Fourier modes $u_\ell$, and then express both  $Q_{2n-1}$ and $I_k$ in terms of $u_\ell$. We develop this method in the appendix \ref{sec:pert} and confirm the expansion \eqref{genQ} up to second order in $I_k$.

\subsection{One-zone potentials}
\label{1cut}
Before we consider one-zone potential in detail, we revisit the zero-zone potential $u=Q_1\equiv 4\lambda_0$ and readily find differential
\bea
dp={d\lambda\over 2\sqrt{\lambda-\lambda_0}}
\eea
to be defined on a Riemann sphere. This is the simplest possible case. 
In this case $p=\sqrt{\lambda-\lambda_0}$,  $u(\varphi)=h=4\lambda_0$ and the whole symplectic space ${\mathcal F}_0(h)$ shrinks to a point. All KdV Hamitonians are fixed  by $h$, $Q_{2n-1}=h^n$ with all action variables identically equal  to zero. 

Next, we consider the differential 
\bea
dp={(\lambda-r)d\lambda\over 2\sqrt{(\lambda-\lambda_0)(\lambda-\lambda_1)(\lambda-\lambda_2)}}
\eea
parameterized by $\lambda_i,r_0$. It is defined on a torus -- a Riemann curve of genus one. We assume that $(\lambda_2,\lambda_1)$ correspond to $k$-th zone. 
After satisfying \eqref{C1} and \eqref{C2}, which requires evaluating elliptic integrals,  we find one-parametric family
\bea
\lambda_2=\lambda_0+{k^2\over 4}\theta_3(\tau)^4,\quad \lambda_1=\lambda_0+{k^2\over 4}\theta_4(\tau)^4, \quad
r=\lambda_0+{k^2\over 4} \theta_4(\tau)^4 \left(1+2 {\partial \ln \theta_3^2(\tau)\over \partial \ln m} \right), \nonumber \\
\label{lambdar}
\eea
where $m=\theta_2^4(\tau)/\theta_3^4(\tau)$ and 
$\tau=i \tau_2$ with positive $\tau_2$. In what follows we use\footnote{Our definition of $q$  is aligned with Wolfram Mathematica. In this section $q$ denotes modular parameter of the genus one elliptic curve $y(\lambda)$. In section \ref{sec:thermalev} we use $q$ to denote modular parameter of the CFT spacetime torus.} $q=e^{i\pi \tau}$  such that $\theta_2=\sum_n q^{(n+1/2)^2}$, $\theta_3=\sum_n q^{n^2}$, $\theta_4=\sum_n (-1)^n q^{n^2}$.

To impose the orbit constraint   $-4p(0)^2=h$ it is more convenient to use the following trick. 
First we evaluate 
\bea
Q_1=4(\lambda_0+\lambda_1+\lambda_2)-8r,
\eea
which expresses $\lambda_0$ in terms of $Q_1$ and $q$ expansion, 
\bea
\label{Q1lambda0}
4\lambda_0=Q_1-k^2\left(\theta_2^4-4\theta_4^4 {\partial \ln \theta_3^2(\tau)\over \partial \ln m}\right) =Q_1-32 k^2 q^2 \left(1+2 q^2+4 q^4 +4 q^6+\dots\right), \qquad
\eea
and then use \eqref{action} to evaluate action variable perturbatively in $q$,
\bea
 I_k &=&\frac{2}{\pi}\int^{\lambda_2}_{\lambda_1} \frac{d\lambda(\lambda-r)\log\lambda}{\sqrt{(\lambda-\lambda_0)(\lambda-\lambda_1)(\lambda_2-\lambda)}}
 \\
    &=&\sum_{n=1}^\infty\frac{2(-1)^n(\lambda_1-\lambda_0)^{n+1}}{n\sqrt{\lambda_2-\lambda_0}\lambda_0^n}
    \left[\frac{F\left(\frac32,\frac12,1;m\right)}{F\left(\frac12,\frac12,1;m\right)}F\left(n+\frac12,\frac12,1;m\right)
    -
    F\left(n+\frac32,\frac12,1;m\right)
    \right]. \nonumber
\eea
Here $F\equiv {}_2F_1$ is the hypergeometric function such that $F\left(\frac32,\frac12,1;m\right)=\theta_3^2$.

An infinite sum over $n$ above has to be evaluated individually for each term in $q$ expansion.
This gives $I_k$ as a function of $\lambda_0$ and $q$, $I_k={32 k^3 q^2\over k^2+4\lambda_0}+O(q^4)$,
which with help of \eqref{Q1lambda0} can be expressed as a function of $Q_1$ and $q$, 
\bea
\label{action_u0-q}
    I_k&=&
    \frac{32 k^3 }{k^2+Q_1}q^2
    +\frac{64 k^3  \left(17 k^4+12 k^2 Q_1+3
   Q_1^2\right)}{\left(k^2+Q_1\right)^3}q^4
   \\
  & &+\frac{128 k^3  \left(5
   k^2+Q_1\right) \left(77 k^6+69 k^4 Q_1+27 k^2 Q_1^2+3
   Q_1^3\right)}{\left(k^2+Q_1\right)^5}q^6
   +\mathcal{O}(q^8). \nonumber
\eea
At this point we use \eqref{q1}, which is exact, $Q_1=h+k I_k. \label{q1z1}$.
Using  $I_k$  given as  a $q$-series expansion with $Q_1$-dependent coefficients  \eqref{action_u0-q},   with help of  \eqref{q1z1} we express $Q_1$ as a series in $q$ with $h$-dependent coefficients by iteratively substituting $Q_1$ written as an $h$-dependent series in $q$. Once we find $Q_1=Q_1(h,q)$, $I_k$ can be deduced from \eqref{q1z1},
\bea
    I_k&=&
    \frac{32 k^3 }{h+k^2}q^2
    +\frac{64 \left(3 h^2 k^3+12 h k^5+k^7\right)}{\left(h+k^2\right)^3}q^4
    \\
    &&+\frac{128 k^3 \left(3 h^4+42 h^3 k^2+108 h^2 k^4-58 h
   k^6+k^8\right)}{\left(h+k^2\right)^5}q^6
   +\mathcal{O}(q^{8}).
\eea
At this point it is straightforward to re-express $q$ as a $h$-dependent power series in $I_k$, $q^2= {h+k^2\over 32 k^3}I_k+O(I_k^2)$.

To obtain coefficients $f^{(n,i)}$  \eqref{genQ} we act as follows. From the definition \eqref{kdveq} we can find $Q_{2n-1}$  as a polynomial in $\lambda_i$ and $r$. 
Using expressions for $\lambda_i, r$ \eqref{lambdar} and \eqref{Q1lambda0}, where $Q_1$ is understood as a function of $h,q$ we write $Q_{2n-1}$ as an $h$-dependent power series in $q$. After that it is straightforward to use $q^2=q^2(h,I_k)$ to re-express $Q_{2n-1}$ as an $h$-dependent power series in $I_k$, 
\begin{align}
\label{genQonecut}
    Q_{2n-1}&=h^n + f^{(n,1)}_k I_k+f^{(n,2)}_{k}I_k^2+f^{(n,3)}_{k}I_k^2
    +\mathcal{O}(I_k^3),
\end{align}
thus fixing $f^{(n,i)}$,
\begin{align}
\label{f1}
    f^{(n,1)}_k&=\sum_{j=0}^{n-1}\frac{\sqrt{\pi}(2n-1)\Gamma(n+1)}{2\Gamma(j+\frac32)\Gamma(n-j)}h^{n-1-j}k^{2j+1}=\sum_{j=0}^{n-1} \xi_n^j\, h^{n-1-j}k^{2j+1},
    \\
\label{f21}
     f^{(n,2)}_k&=\sum_{j=0}^{n-1}
      \frac{\sqrt{\pi}(2n-1)\Gamma(n+1)(j(2n+1)-2n+2)}{16\,\Gamma\left(j+\frac{3}{2}\right)\Gamma(n-j)}h^{n-1-j}k^{2j},    \\
\label{f31}
f^{(n,3)}_k&={(2n-1)n(n-1)\over 64 k^3}h^n+\\ \nonumber
& \quad  \sum_{j=0}^{n-1} 
      \frac{\sqrt{\pi}(2n-1)\Gamma(n+1)\, {\sf p}}{1536\,\Gamma\left(j+\frac{5}{2}\right)\Gamma(n-j)}h^{n-1-j}k^{2j-1},\\ \nonumber
{\sf p}&=4 j^3(2n+1)(2n+3)-2j^2(2n+1)(10n-21)-3j(2n+3)(10n-7)+36(n-1)(2n-1).
\end{align}
More technical details concering one-zone calculation can be found in appendix \ref{A:1cut}.

\subsection{Two-zone potentials}
\label{2cut}
In case of  two zones  the differential 
\bea
dp={(\lambda-r_1)(\lambda-r_2)d\lambda\over 2\sqrt{\prod_{i=0}^4(\lambda-\lambda_i)}}
\eea
depends on seven parameters subject to 4 constraints \eqref{C1} and \eqref{C2}. Corresponding integrals can not be evaluated analytically. We therefore proceed by expanding perturbatively, assuming both zones, and hence corresponding action variables, are small.  
We introduce two infinitesimal variables $\epsilon_1,\epsilon_2$ of the same order, such that $\lambda_2-\lambda_1$ is of order $\epsilon_1$ and $\lambda_4-\lambda_3$ is of order $\epsilon_2$. Action variables are quadratic in $\epsilon_i$, $I_k \sim \epsilon_1^2, I_\ell \sim \epsilon_2^2$, where we assumed  $(\lambda_1,\lambda_2)$ and $(\lambda_3,\lambda_4)$  correspond to $k$-th and $\ell$-zones respectively. Our goal is to find $Q_{2n-1}$ up to third order in the pertubative expansion in $I_k,I_\ell$. Hence in what follows we must expand all quantities in $\epsilon_i$ up to sixth order. The details of this calculation can be found in  in Appendix \ref{perturbative}.

After satisfying  \eqref{C1} and \eqref{C2} we find $\lambda_i$ for $i\geq 1$ and $r_i$ in terms of $\lambda_0$ and $\epsilon_1,\epsilon_2$, as a perturbative expansion in $\epsilon_i$. 
Then, we evaluate $I_k,h$ and $Q_{2n-1}$ also as function of $\lambda_0$ and $\epsilon_1,\epsilon_2$, similarly expanding in $\epsilon_i$ up to and including sixth order. 
By matching both sides of \eqref{genQ} we find coefficients $f^{(m,n)}_{k,\ell}$, yielding 
\bea
\label{f22}
f^{(n,2)}_{k,\ell}&=& \sum_{j=1}^{n-1}\frac{\sqrt{\pi}(2n-1)^2\Gamma(n+1)}{4 \Gamma(n-j) \Gamma\left(j+\frac{3}{2}\right)} h^{n-1-j}
    \sum_{s=0}^{j-1}k^{2(j-s)-1}\ell^{2s+1},\\
\label{f3}
f^{(n,3)}_{k,\ell}&=& {\ell\over (k^2-\ell^2)^2} \left(-{(2n-1)n(n-1)\over 4}h^n +
\sum_{j=0}^{n-1} {\sqrt{\pi}(2n-1)^2\Gamma(n+1)\over 64 \Gamma(n-j) \Gamma\left(j+\frac{5}{2}\right)}h^{n-1-j}\, {\sf q}\right), \\ \nonumber 
{\sf q}&=&-4(2n+1){k^{2j+4}-\ell^{2j+4}\over k^2-\ell^2}+k^{2j+2}(3+2j)(j(2n+1)-4n+5)+\ell^{2j+2}2(3j+n+5)+ \\
\qquad &&k^{2j}\ell^2 (3+2j)(j(2n+1)-2n+2)+k^2 \ell^{2j}(3+2j)(4n-1). \nonumber
\eea

\subsection{Three-zone potentials}
\label{3cut}
Extending calculations of the previous section using the technique of appendix \ref{perturbative} to the three-zone case we can fix 
\bea
f^{(n,3)}_{k,\ell,p}= \sum_{j=0}^{n-3}{\sqrt{\pi}(2n-1)^3\Gamma(n+1)(n-2-j)\over 8 \Gamma(n-1-j) \Gamma\left(j+{7\over 2}\right)}h^{n-3-j} \sum_{s_1=0}^{j} \sum_{s_2=0}^{j-s_1} k^{2j+1-2(s_1+s_2)} \ell^{2s_1+1} p^{2s_2+1}.  \nonumber \\ \label{f33}
\eea

\subsection{Consistency check}
In case of an $m$-zone potential we can parametrize the differential $dp$ with help of $\lambda_0$ and $\epsilon_i$, $1\leq i\leq m$, cf.~(\ref{l1}-\ref{r2}),
\bea
\label{lr1}
\lambda_i&=&\lambda_0+\dots, \qquad 1\leq i \leq 2m,\\
r_i&=&\lambda_0+\dots, \qquad 1\leq i \leq m, \label{lr2}
\eea
where dots stand for $\epsilon_i$ but not $\lambda_0$-dependent terms. Similarly action variables $I_k$, charges $Q_{2n-1}$ and the orbit parameter $h=-4p(0)^2$ will be some functions of $\lambda_0$ and $\epsilon_i$. While dependence of $I_k$ and $h$ on $\lambda_0$ is non-trivial, since $Q_{2n-1}$ are the coefficients of $1/\lambda$ expansion of $p(\lambda)$ at infinity and $\lambda_0$ is simply the shift of the argument of $p(\lambda)$, we find
\begin{align}
\label{relationh=0}
    Q_{2n-1}=\sum_{k=0}^{n} \frac{\Gamma(n+1)}{\Gamma(k+1)\Gamma(n-k+1)}(4\lambda_0)^{n-k}Q_{2k-1}^0.
\end{align}
Here $Q_{2k-1}^0$ are the charges evaluated with help of \eqref{kdveq} taking $\lambda_0=0$ in (\ref{lr1},\ref{lr2}).
Assuming we know $Q_{2n-1}(h,I_k)$ where $h=h(\lambda_0,\epsilon_i)$ and $I_k=I_k(\lambda_0,\epsilon_i)$,
one can introduce $I_k^0=I_k(0,\epsilon_i)$ such that $Q_{2k-1}^0=Q_{2k-1}(0,I_k^0)$. Here first argument is zero simply because $h(0,\epsilon_i)=0$. Then both sides of equation \eqref{relationh=0} become functions of $\lambda_0$ and $\epsilon_i$, providing a non-trivial check. 

There is an alternative way to use \eqref{relationh=0} to check the consistency of  the perturbative expansion\eqref{genQ} with the coefficients found in the text. We can invert $h=h(\lambda_0,\epsilon_i)$ and $I_k=I_k(\lambda_0,\epsilon_i)$  to express both $\lambda_0$ and $I_k^0$ via $h$ and $I_k$, 
\bea
\nonumber
\lambda_0&=&h+\sum_k -{h I_k\over k}+\frac{h \left(h+5 k^2\right)I_k^2}{8 k^4}-\frac{h \left(5 h^2+30 h k^2+41 k^4\right)I_k^3}{128 k^7}+
\sum_{k<\ell} {h I_k I_\ell\over k \ell} + \\ 
&&\sum_{k\neq \ell}\frac{h I_k^2 I_\ell \left(h^2 \ell^2-h \left(k^4-4 k^2 \ell^2+\ell^4\right)-5 k^6+11 k^4 l^2-5 k^2 \ell^4\right)}{8 k^4 \ell (k-\ell)^2 (k+\ell)^2}
-\sum_{k<\ell<p}\frac{h I_k I_\ell I_p}{k \ell p}+{\mathcal O}(I^4),  \nonumber\\ \nonumber
I_k^0&=&I_k+{h I_k\over k^2}-\frac{h I_k^2 \left(h+5 k^2\right)}{8 k^5}+\frac{h I_k^3 \left(5 h^2+30 h k^2+41 k^4\right)}{128 k^8}-\sum_{\ell\neq k}\frac{h I_k I_\ell \left(h+k^2\right)}{k^2 \ell \left(k^2-\ell^2\right)}+\\ \nonumber
&& \sum_{\ell \neq k}\frac{h I_k^2 I_\ell \left(2 h^2 \left(-k^4+2 k^2 \ell^2+\ell^4\right)+h k^2 \left(7 k^4-14 k^2 \ell^2+15 \ell^4\right)+k^4 \left(5 k^4-10 k^2 \ell^2+9 \ell^4\right)\right)}{8 k^5 l \left(k^2-\ell^2\right)^3}+ \\ \nonumber
&&  \sum_{\ell \neq k} \frac{h I_k I_\ell^2 \left(h^2 \left(k^4+5 k^2 \ell^2-2 \ell^4\right)+h \left(k^6+10 k^4 \ell^2-9 k^2 \ell^4+6 l^6\right)+k^2 \ell^2 \left(5 k^4-7 k^2 \ell^2+6 l^4\right)\right)}{8 k^2 \ell^4 \left(k^2-\ell^2\right)^3} + \\ \nonumber
&& \sum_{p\neq \ell \neq k} \frac{h I_k I_\ell I_p \left(2 h^2+3 h k^2+k^4\right)}{k^2 \ell p \left(k^2-\ell^2\right) \left(k^2-p^2\right)}+{\mathcal O}(I^4).
\eea
Now $Q_{2n-1}^0(0,I_k^0(h,I_k))$ is a function of $h, I_k$ and \eqref{relationh=0} provides a non-trivial check for the coefficients in \eqref{genQ}.

This check also ensures that $Q_{2n-1}(h,I_k)$ satisfy another identity
\bea
\label{check2}
{1\over n+1}{\partial Q_{2n+1}\over \partial u_0}= Q_{2n-1},
\eea 
which follows from the properties of Gelfand-Dikii polynomials \eqref{iterativeR}. Here $Q_{2n-1}[u(\varphi)]$ are understood as functionals of $u(\varphi)$ and the derivative is with respect the zero Fourier mode of $u(\varphi)$, while all other Fourier modes are kept fixed. The shift of $u_0$ with all other modes intact is equivalent to a shift of the spectrum by a constant, hence 
\bea
\left({\partial \over \partial u_0}\right)_{u_\ell}=4\left({\partial \over \partial \lambda_0}\right)_{\epsilon_i}.
\eea
Then \eqref{check2} follows immediately from the right-hand-side of \eqref{relationh=0}.

For an $m$-zone potential, all higher KdV charges $Q_{2n-1}$ are some functions of first $m+1$ charges. Thus for one-zone potentials $Q_5, Q_7,\dots$ are functions of $Q_1,Q_3$, see e.g.~section 2.4 of \cite{Dymarsky_2020} for details. For the three-zone potentials higher $Q_{2n-1}$ would depend on $Q_1,Q_3,Q_5,Q_7$ 
In principle this provides additional consistency check for \eqref{genQ}. In practice the dependence is so complicated, it doesn't provide a useful check even for the one-zone case. 

%

\section{``Energies'' of primary states via ODE/IM correspondence}
\label{sec:vacuumenergy}
In the previous section we found classical expression for $Q_{2n-1}$ in term of action variables $I_k$ and the orbit invariant $h$. Following the standard rules of semiclassical quantization $I_k$ should be promoted to an integer quantum  number, while $h$ will become the dimension of the highest  weight (primary) state $\Delta$, marking representation of the Virasoro algebra. It is easy to see, this naive receipt fails already for the values of $Q_{2n-1}$ on a primary state $|\Delta\rangle$. Indeed, taking all $I_k$ to zero, we  readily find $Q_{2n-1}=h^k$, which upon the  naive quantization yields ${\sf Q}^0_{2n-1}=\Delta^n$ where 
\bea
Q_{2n-1}|\Delta\rangle={\sf Q}^0_{2n-1}|\Delta\rangle.
\eea
This answer is missing $c$-dependent terms. Explicit values of ${\sf Q}^0_{2n-1}$ for $n\leq 8$ were calculated in \cite{bazhanov97zero} via brute-force approach, using explicit expressions for $Q_{2n-1}$ in terms of free field representation. The pattern is clear, while $\Delta^n$ is indeed the leading term, full expression is a polynomial in both $\Delta$ and $c$ of order $n$.

There is no known receipt to obtain exact ${\sf Q}^0_{2n-1}$  from the semiclassical quantization, hence our strategy will be the following. We will combine exact expression for ${\sf Q}^0_{2n-1}$ in the large $c$ limit, which will be obtained in this section by a different method, with the classical result of section \ref{sec:classic}, to find spectrum of  excited states in the large $c$ limit in next section. 

To find ${\sf Q}^0_{2n-1}$ we use ODE/IM correspondence, initiated in \cite{bazhanov2001spectral,bazhanov2003higher} and more recently developed  in \cite{dorey2020geometric} (also see \cite{Conti:2021xzr}), which relates qKdV spectrum  to solutions of an auxiliary Schr$\ddot{\rm o}$dinger equation
\begin{align}
\label{Schrodinger}
    \partial_x^2 \Psi(x)+\left(E-x^{2\alpha}-\frac{l(l+1)}{x^2}\right)\Psi(x)=0,
\end{align}
where 
\begin{align}
(l+1/2)^2=4(\alpha+1)\tilde{\Delta}, \quad \tilde{c}=-\frac{\alpha^2}{4(\alpha+1)}.
\end{align} 
Equation \eqref{Schrodinger} can be solved using WKB approximation by systematically expanding in a small parameter. This leads to a quadratic ODE which can be solved iteratively. We delegate all details to Appendix \ref{qvac} and only write down  iterative relation which defines coefficients $c^{(n)}_k$ for   $n\geq 1$, $n\geq k \geq 0$,
\begin{align}
\label{ckn1}
   \sum_{j=0}^{n}\sum_{p=0}^j\sum_{q=0}^{n-j}\delta_{p+q,k} c_p^{(j)}c_q^{(n-j)}
     - 2\left[n-k-u-\frac{n-2}{2\alpha}\right]c_{k-1}^{(n-1)}
  +
  \left(2k-3n+4\right)c_k^{(n-1)}=0,
\end{align}
and we formally assumed $c^{(n)}_{-1}=c^{(n)}_{n+1}=0$, $u^2=-\tilde{\Delta}/\tilde{c}$, and  the starting values are 
\begin{align}
\label{ckn0}
    c_0^{(0)}=-\frac{1}{\alpha},\qquad c_0^{(1)}=-\frac{1}{2}, \quad c_1^{(1)}=\frac{1}{2\alpha}-u.
\end{align}
Coefficients $c_k^{(n)}$ determine values of $Q_{2n-1}$ acting on primaries \cite{dorey2020geometric}, 
\begin{align}
{\sf Q}^0_{2n-1}
   =\frac{(2n-1)\Gamma(n+1)}{\sqrt{\pi}\Gamma(1-\frac{2n-1}{2\alpha})4^n(\alpha+1)^n}
    \sum_{k=0}^{2n}c_k^{(2n)}\Gamma\left(k+\frac{3}{2}-3n\right)\Gamma\left(2n-k-\frac{2n-1}{2\alpha}\right). \label{vacenergy}
\end{align}
Although this is not obvious, ${\sf Q}^0_{2n-1}$ given by \eqref{vacenergy} is a polynomial in terms of $\tilde{\Delta}$ and $\tilde{c}$. After some algebra we find leading order expansion
\begin{align}
\label{vacuumE}
     {\sf Q}^0_{2n-1}&=\tilde{\Delta}^n
    +\sum_{j=0}^{n-1}
   \tilde R^{(1)}_{n,j}\tilde{\Delta}^{n-j-1}\, \tilde{c}^j
    +\sum_{j=0}^{n-2}
   \tilde R^{(2)}_{n,j}\tilde{\Delta}^{n-j-2}\, {\tilde c}^{j}+\sum_{j=0}^{n-3}
   \tilde R^{(3)}_{n,j}\tilde{\Delta}^{n-j-3}\, {\tilde c}^{j}+\mathcal{O}({\tilde c}^{n-3}).
\end{align}
where 
\begin{align}
\label{R1}
    \tilde R^{(1)}_{n,j}&=\frac{(2n-1)\sqrt{\pi}\Gamma(n+1)}{4\Gamma(j+\frac32)\Gamma(n-j)} \zeta(-2j-1)=\xi_n^j {\zeta(-2j-1)\over 2},
     \\
\label{R2}
\tilde{R}^{(2)}_{n,j}&=\frac{(2n-1)\sqrt{\pi}\Gamma(n+1)}{24\times 4\Gamma(j+\frac52)\Gamma(n-j-1)}\times 
\\ \nonumber 
&    \Biggl\{-6\zeta(-2j-3)\left(2j+3-(2n-1)y_1(j+1)\right)
    +3(2n-1)\zeta_2(j)\, 
    \Biggr\},   \nonumber \\
\tilde{R}^{(3)}_{n,j}&=\frac{(2n-1)\sqrt{\pi}\Gamma(n+1)}{24^2\times 4\Gamma(j+\frac72)\Gamma(n-j-2)}
\Biggl\{
6^2 \zeta(-2j-5)(2 j^2 + 7 j + 5)-  
(2n-1)r_{n,j}
\Biggr\},\nonumber\\ \nonumber 
{r}_{n,j}&=12 \zeta_3(j)+ 36 \zeta_2(j+1) (y_1(j+2)+j+2)+3 \left(4 j^2+18 j+23\right) \zeta (-2 j-3)+\\ &36 \zeta (-2 j-5) \left(y_1^2(j+2)+2 (j+2) y_1(j+2)+y_2(j+2)\right)+ (2 n+1)p_j. \label{R3}
\end{align}
Functions $\zeta_2,\zeta_3, y_1,y_3$  are defined in the Appendix \ref{qvac}, where we also give values of $p_j$ for $0\leq j\leq 17$.

\section{Spectrum of quantum $Q_{2k-1}$}
\label{sec:quantization}
At this point we are ready to combine classical pertubative expression for $Q_{2n-1}(h,I_k)$ \eqref{genQ} with the ``energies'' of primary state \eqref{vacuumE} to obtain ${\sf Q}_{2n-1}$ up to first two non-trivial orders in $1/\tilde{c}$ expansion. 

The naive semi-classical quantization  would map  the co-adjoint orbit invariant $h$ and the actions variables $I_k$ on the classical side to dimension of the primary state $\Delta$ and the excited state quantum numbers $n_k$ correspondingly,
\bea
h\rightarrow {24\Delta\over c}, \qquad I_k \rightarrow {24 n_k\over c}. \label{qnaive}
\eea 
Also classical charge $Q_{2n-1}$ should be rescaled by $(c/24)^n$.  Starting from \eqref{genQ} this correctly reproduces full quantum spectrum of $Q_1$ and the leading $\Delta^n$ term in $Q_{2n-1}$. But it falls short of reproducing sub-leading terms even for the primary state \eqref{vacuumE}. The relation between classical and quantum quantities \eqref{qnaive} is only correct at the leading $c$ order. In \cite{GGE2} we observed that using $c-1$  as an expansion parameter leads to more elegant expressions. 
This is confirmed by \eqref{vacuumE}, which looks most naturally if written in terms of $\tilde{\Delta}$ and $\tilde{c}$.
We therefore propose the following quantization map, which agrees with the naive one at leading order, 
\bea
h\rightarrow {\tilde{\Delta}\over \tilde{c}}, \qquad I_k \rightarrow {n_k\over \tilde{c}}, \qquad \tilde{\Delta}=\Delta-\tilde{c},\qquad \tilde{c}={c-1\over 24}.  \label{quantization}
\eea  
This does not solve the problem of reproducing subleadig terms in ${\sf Q}_{2n-1}^0$, but this can be fixed, at least at first subleading  order, by introducing the Maslov index, $n_k\rightarrow \tilde{n}_k=n_k+1/2$.  
We thus arrive at the following map, 
\bea
Q_{2n-1}(h,I_k) \rightarrow  {\sf Q}_{2n-1}=\tilde{c}^n\, Q_{2n-1}(\tilde{\Delta}/\tilde{c},(n_k+1/2)/\tilde{c}). \label{Quant}
\eea
Infinite sums due to Maslov index contributing to ``vacuum energy''  should be regularized using zeta-function regularization. It is now straightforward to see that we immediately reproduce the leading $1/\tilde{c}$ term \eqref{R1},
\bea
\nonumber
&&Q_{2n-1}=h^n+\sum_k f_k^{(n,1)}(h)\, I_k +O(I^2) \rightarrow {\sf Q}_{2n-1}= \tilde{\Delta}^n+\tilde{c}^{n-1} \sum_k f_k^{(n,1)}(\tilde{\Delta}/\tilde{c})\, \tilde{n}_k +O(\tilde{c}^{n-2}) =\\ \nonumber
&&\tilde{\Delta}^n+ \sum_k \sum_{j=0}^{n-1} \xi_n^j\, \tilde{\Delta}^{n-1-j} \tilde{c}^{j}  k^{2j+1}  (n_k+1/2)+O(\tilde{c}^{n-2}) =\\
&&\tilde{\Delta}^n+ \sum_{j=0}^{n-1} \xi_n^j\, \tilde{\Delta}^{n-1-j} \tilde{c}^{j}   \left(\sum_k k^{2j+1} n_k+{\zeta(-2j-1)\over 2}\right)+O(\tilde{c}^{n-2})
\label{lQ}
\eea 
In other words, at first sub-leading order $\tilde{c}^{n-1}$ the quantization prescription \eqref{Quant}
leads to \eqref{leadingQ} which passes all available tests: matches the spectrum of $Q_1,Q_3,Q_5,Q_7$ (see section \ref{sec:check} below) and thermal expectation values for $Q_9,\dots, Q_{13}$ (see section \ref{sec:thermalev} below) at the order $\tilde{c}^{n-1}$.

There is another way to write \eqref{lQ}. We can express ${\sf Q}_{2n-1}$ as ${\sf Q}_{2n-1}^0$ plus the terms from the classical $Q_{2n-1}$ \eqref{genQ} which non-trivially depend on $I_k$ using the substitution \eqref{quantization}, i.e.~without the Maslov index, 
\bea
\label{1l}
{\sf Q}_{2n-1}= {\sf Q}_{2n-1}^0 +\tilde{c}^{n-1} \sum_k f_k^{(n,1)}(\tilde{\Delta}/\tilde{c})\, n_k +O(\tilde{c}^{n-2}).
\eea
At $\tilde{c}^{n-1}$ order it is the same as \eqref{lQ}. 

To obtain the quantum spectrum at next  order $\tilde{c}^{n-2}$,  we could try the prescription \eqref{Quant}, apply the zeta-function regularization and notice that many but not all terms from \eqref{R2} are reproduced. Thus, we see that the quantization \eqref{Quant} is exact only at leading $1/\tilde{c}$ order, at higher orders  the expression obtained from the classical $Q_{2n-1}$ has to be modified as well. Indeed, starting from the classical \eqref{genQ} and using substitution \eqref{quantization} we would find that terms contributing at the order $\tilde{c}^{n-p}$ are homogeneous polynomials in $n_k$ of order $p$. This is very restrictive and obviously incorrect. We already saw that even at the first sub-leading order $\tilde{c}^{n-1}$ the homogeneous (linear) in $n_k$ terms have to be amended by a constant, i.e.~$(n_k)^0$ term. This suggest the following ``quantization rules'': to obtain the quantum spectrum ${\sf Q}_{2n-1}$ in $1/\tilde{c}$ expansion, one starts with the classical perturbation expression \eqref{genQ} and make the substitution \eqref{quantization}, together with the overall rescaling by $\tilde{c}^n$.  As the order $\tilde{c}^{n-p}$ this fixes leading, homogeneous in $n_k$ terms of order  $p$. These terms should be amended by the sub-leading terms of order $p-1$, $p-2$, \dots, $0$ in $n_k$. These terms should be regarded as quantum corrections and  should be determined separately, they do not follow from the classical answer in any simple way. More explicitly, 
\bea
\nonumber
&&{\sf Q}_{2n-1}=\tilde{\Delta}^n +\tilde{c}^{n-1}\left(\sum_k g^{(1)}_{k} n_k+ g^{(1)} \right) +
\tilde{c}^{n-2}\left(\sum_{k_1,k_2} g^{(2)}_{k_1,k_2} n_{k_1} n_{k_2}+ \sum_k g^{(2)}_{k} n_k+g^{(2)} \right)+\\
&&\tilde{c}^{n-3}\left(\sum_{k_1,k_2,k_3} g^{(3)}_{k_1,k_2,k_3} n_{k_1} n_{k_2} n_{k_3}+\sum_{k_1,k_2} g^{(3)}_{k_1,k_2} n_{k_1} n_{k_2}+ \sum_k g^{(3)}_{k} n_k+g^{(3)} \right)+\dots
\label{classrep}
\eea
Here $g^{(p)}$ with different number of indexes denote different quantities. The leading terms $g^{(p)}_{k_1,\dots, k_p}$ are given by classical expressions \eqref{genQ} upon the substitution \eqref{quantization}
\bea
g^{(1)}_k&=&f^{(n,1)}_k(\tilde{\Delta}/\tilde{c}),\\
g^{(2)}_{k\ell}&=&{1\over 2}f^{(n,2)}_{k\ell},  \quad g^{(2)}_{kk}=f^{(n,2)}_k,\\
g^{(3)}_{k\ell m}&=&{1\over 6}f^{(n,3)}_{k\ell m},  \quad g^{(3)}_{k k \ell}={1\over 3}f^{(n,3)}_{k \ell},\quad 
g^{(3)}_{k k k}=f^{(n,3)}_k,
\eea
for $k\neq \ell \neq m$ and $g^{(p)}$ are given by  (\ref{R1},\ref{R2},\ref{R3}). 
This is essentially the generalization of \eqref{1l} to higher orders in $1/\tilde{c}$. 
Coefficients $g^{(2)}_k$,  $g^{(3)}_{k\ell},g^{(3)}_{k}$, etc.~are quantum corrections and a priory not known. 

To fix $g^{(2)}_k$ we employ the following strategy,  we will try to ``salvage'' the Maslov index quantization \eqref{Quant} by adding minimal possible terms subleading in powers of $n_k$, 
\bea
\nonumber
{\sf Q}_{2n-1}=\tilde{\Delta}^n +\tilde{c}^{n-1} \sum_k g^{(1)}_{k} \tilde{n}_k +
\tilde{c}^{n-2}\left(\sum_{k_1,k_2} g^{(2)}_{k_1,k_2} \tilde{n}_{k_1}\tilde{n}_{k_2}+ \sum_k \tilde{g}^{(2)}_{k} \tilde{n}_k+\tilde{g}^{(2)} \right)+\dots\\
\label{2l}
\eea
This expression is understood in terms of the zeta-function regularization and $\tilde{g}^{(2)}_{k}, \tilde{g}^{(2)}$ are different from $g^{(2)}_{k},g^{(2)}$. Our goal is to reproduce ``vacuum energy'' ${\sf Q}_{2n-1}^0$. There is infinitely many ways to do that, for example by taking $g^{(2)}_{k}=0$, $\tilde{g}^{(2)}=g^{(2)}$, but we will additionally require that the zeta-functions from \eqref{R2} will become the sums of the form $\sum_k k^p$ in \eqref{2l}. This leads to 
\bea
\label{tg2}
\tilde{g}^{(2)}_k=\sum_{j=0}^{n-1}{1\over 4}\xi_n^j \left((2n-1)y_1(j)-2 j-1\right) \tilde{\Delta}^{n-1-j}\tilde{c}^j k^{2j+1},
\eea
and very simple 
\bea
\tilde{g}^{(2)}=  -\frac{n(n-1)(2n-1)\tilde{\Delta}^{n-1}}{96 \tilde{c}}.
\eea
This term is necessary to subtract $n_k$-independent $\tilde{\Delta}^{n-1}\tilde{c}^{-1}$ term coming  from $\sum_k \tilde{g}^{(2)}_k \tilde{n}_k$ to match  ${\sf Q}_{2n-1}^0$ \eqref{vacuumE} which has no terms with the negative powers of $c$.

For convenience we give the full expression \eqref{2l} explicitly 
\begin{align}
\label{resultpreliminary}
{\sf Q}_{2n-1}&=\tilde{\Delta}^n
    +\sum_k\sum_{j=0}^{n-1}\frac{(2n-1)\sqrt{\pi}\,\Gamma(n+1)}{2\Gamma(j+\frac32)\Gamma(n-j)}\tilde{\Delta}^{n-1-j}\tilde{c}^j k^{2j+1}\tilde{n}_k
    \\  \nonumber
    &\quad
    -\frac{n(n-1)(2n-1)\tilde{\Delta}^{n-1}}{96 \tilde{c}}
    \\  \nonumber
    &\quad
    +\sum_k\sum_{j=0}^{n-1}\frac{(2n-1)\sqrt{\pi}\,\Gamma(n+1)}{8\Gamma(j+\frac32)\Gamma(n-j)}\left((2n-1)y_1(j)-2 j-1\right)\tilde{\Delta}^{n-1-j} \tilde{c}^{j-1} k^{2j+1}\tilde{n}_k
    \\ \nonumber
    &\quad-\sum_k \sum_{j=0}^{n-1}\frac{(2n-1)\sqrt{\pi}\, \Gamma(n+1)(2n j+2n-3j-2)}{16\,\Gamma\left(j+\frac{3}{2}\right)\Gamma(n-j)}\tilde{\Delta}^{n-j-1} \tilde{c}^{j-1} k^{2j}\tilde{n}_k^2
       \\  \nonumber
       &\quad+
    \frac{1}{2}\sum_{k,\ell} \sum_{j=1}^{n-1}\frac{(2n-1)^2 \sqrt{\pi}\, \Gamma(n+1)  }{4 \Gamma\left(j+\frac{3}{2}\right) \Gamma(n-j)}\tilde{\Delta}^{n-j-1}\tilde{c}^{j-1}
    \sum_{s=0}^{j-1}k^{2(j-s)-1}\ell^{2s+1}\tilde{n}_k
    \tilde{n}_\ell
    +\mathcal{O}(c^{n-3}). \nonumber
\end{align}
We conjecture this is the full quantum spectrum of $Q_{2n-1}$ up to $\tilde{c}^{n -2}$ order and verify that  it passes all available checks. 

From here it is now straightforward to find ${\sf Q}_{2n-1}$ in the representation \eqref{classrep}. Coefficient 
\bea
&&g^{(2)}_{k}= \sum_{j=0}^{n-1}\frac{(2n-1) \sqrt{\pi}\, \Gamma(n+1)  }{8 \Gamma\left(j+\frac{3}{2}\right) \Gamma(n-j)} v(n,j,k) \tilde{\Delta}^{n-j-1}\tilde{c}^{j-1},   \\
\nonumber
&&
v(n,j,k)=\\
&& (2n-1)\sum_{s=0}^{j-1}\zeta(2(s-j)+1)k^{2s+1} +((2n-1)y_1(j)-2 j-1) k^{2j+1} -{1\over 2}(2n j+2n-3j-2) k^{2j} \nonumber
\eea
is significantly more bulky than \eqref{tg2}, while  the full expression is 
\bea
\label{result}
{\sf Q}_{2n-1}&=& {\sf Q}_{2n-1}^0+ \sum_k\sum_{j=0}^{n-1}\xi_n^j \tilde{\Delta}^{n-j-1}\tilde{c}^j k^{2j+1}n_k \\ \nonumber
&+&\sum_{k,\ell} \sum_{j=1}^{n-1} \xi_n^j {(2n-1)\over 4} \tilde{\Delta}^{n-j-1}\tilde{c}^{j-1}
    \sum_{s=0}^{j-1}k^{2(j-s)-1}\ell^{2s+1} n_k
    n_\ell \\  	\nonumber
&-&\sum_k \sum_{j=0}^{n-1}\xi_n^j {(2n j+2n-3j-2)\over 8} \tilde{\Delta}^{n-j-1} \tilde{c}^{j-1} k^{2j}n_k^2
       \\  \nonumber
&+&  \sum_k\sum_{j=0}^{n-1} \xi_n^j  {v(n,j,k)\over 4} \tilde{\Delta}^{n-j-1}\tilde{c}^{j-1} n_k
    +\mathcal{O}(c^{n-3}). \nonumber
\eea

To summarize, we have found the (conjectured) spectrum of all qKdV charges at first two sub-leading orders in $1/c$ expansion (\ref{resultpreliminary},\ref{result}) and observed certain patterns which may help fix the spectrum at higher orders. Let us spell the step to find the next $1/\tilde{c}^3$ order, i.e.~fix the terms of order $\tilde{c}^{n-3}$ in \eqref{classrep}. The classical result for $Q_{2n-1}$ in terms of action variables $I_k$ was calculated up to cubic order in (\ref{f31},\ref{f3},\ref{f33}).  ``Energies'' of primary states ${\sf Q}_{2n-1}^0$ were also calculated to this order, see eq.~\eqref{R3}. Thus $g^{(3)}_{k_1 k_2 k_3}$ and $g^{(3)}$ are known, and  to find the spectrum one would only need to fix $g^{(3)}_{k_1 k_2}$ and $g^{(3)}_{k}$. To do that one would need to find $\tilde{g}^{(3)}_{k_1 k_2}$ and $\tilde{g}^{(3)}_{k}$  from the expansion \eqref{2l} to reproduce \eqref{R3} via zeta-function regularization and  minimal possible $\tilde{g}^{(3)}$, which presumably will only include terms with negative powers of $\tilde{c}$. ``Restoring'' 
$\tilde{g}^{(3)}_{k_1 k_2}$ and $\tilde{g}^{(3)}_{k}$ from $\tilde{R}_{n,j}^{(3)}$ is not a mathematically  well-posed problem. We expect that  all zeta-functions $\zeta(-2j-1)$ in $\tilde{R}_{n,j}^{(3)}$ to lead to the sums $\sum_k k^{2j+1} \tilde{n}_k$ -- the rule which successfully worked at second $1/\tilde{c}$ order.  At third order this rule should be amended by others, as suggested by a non-polynomial dependence on $k$ in \eqref{f3}. In practice, restoring  $\tilde{g}^{(3)}_{k}$ from $\tilde{R}_{n,j}^{(3)}$ may require establishing the analytic form of coefficients $p_j$ in \eqref{R3} and then reverse-engineering corresponding $k_1,k_2,k_3$-dependent sums. Once hypothetical  $\tilde{g}^{(3)}_{k_1 k_2}$ and $\tilde{g}^{(3)}_{k}$, and accordingly ${g}^{(3)}_{k_1 k_2}$ and ${g}^{(3)}_{k}$ are fixed, a non-trivial set of checks  is provided by the spectrum of $Q_3,Q_5,Q_7$ generated by computer algebra, as well as the requirement that thermal expectation values $\langle Q_{2n-1}\rangle_q$ discussed in section \ref{sec:thermalev} must have certain modular properties.

\subsection{Computer algebra check}
\label{sec:check}
For $n=1$ the expansion \eqref{result} reduces to \eqref{Q1} which is a simple check. A more sophisticated check is provided by $Q_3, Q_5$ and $Q_7$ which are known explicitly  in terms of the Virasoro algebra generators \cite{bazhanov1996integrable} 
\bea
Q_3&=&\left(L_0^2-{c+2\over 12}L_0+{c(5c+22)\over 2990}\right)+\tilde{Q}_3,\\ \nonumber
\tilde{Q}_3&=&2\sum_{k=1}^\infty L_{-k}L_k,\\
Q_5&=&\left(L_0^3- \frac{c+4}{8} L_0^2 + \frac{(c+2)(3c + 20)}{576}L_0  - \frac{c(3c + 14)(7c + 68)}{290304}\right)+ \tilde{Q}_5, \\
\tilde{Q}_5 &=& \sum_{k, l = 0}^{\infty} L_{-k-l} L_k L_l + 2\sum_{k = 1, l = 0}^{\infty} L_{-k} L_{k-l} L_l + \sum_{k, l = 1}^{\infty} L_{-k} L_{-l} L_{k+l} \, + \nonumber  \\ &+& \sum_{n=1}^{\infty} \left(\frac{c+2}{6}m^2 -\frac{c}{4} - 1 \right) L_{-n}L_n - L_0^3,
\eea
and  \cite{dymarsky2019zero}
\bea
\nonumber
Q_7 &=& \sum_{k, l, m = 1}^{\infty} L_{-k} L_{-l} L_{-m} L_{k+l+m} + \sum_{k, l, m = 0}^{\infty} L_{-k-l-m} L_{k} L_{l} L_{m} + \\ && 3\sum_{\substack{k, l = 1  \\ m= 0}}^{\infty} L_{-k} L_{-l} L_{k+l-m} L_m + 3 \sum_{\substack{k = 1  \\ l, m= 0}}^{\infty} L_{-k} L_{k-l-m} L_l L_m + \nonumber \\  && \frac{8+c}{3} \left[\sum_{k, l =1}^{\infty} (k+l)l L_{-k} L_{-l}L_{k+l} + \sum_{\substack{k = 1  \\ l= 0}}^{\infty} (k-l)k L_{-k} L_{k-l} L_l \right] + \nonumber  \\  \nonumber && \frac{8+c}{3} \left[\sum_{k, l =0}^{\infty} (k+l)k L_{-k-l} L_{k}L_{l} + \sum_{\substack{k = 0  \\ l= 1}}^{\infty} (k-l)k L_{-l} L_{l-k} L_k \right] + \\ && \sum_{n=1}^{\infty} \left(\frac{c^2 - c -141}{90}n^4 - \frac{7c + 59}{18}n^2 \right) L_{-n}L_n - \left(\frac{1}{48}c^2 +\frac{53}{360} c + \frac{19}{90} \right) \tilde{Q}_3 - \nonumber \\  \nonumber && \left(\frac{1}{6} c + 1 \right) \tilde{Q}_5  - \frac{c+6}{6} L^3_0 + \frac{15c^2 + 194c + 568}{1440}L_0^2 - \\ && \frac{(c+2)(c+10)(3c+28)}{10368} L_0  + \frac{c(3c+46)(25c^2+ 426c + 1400)}{24883200}. 
\eea
Using computer algebra spectrum of $Q_3,Q_5,Q_7$ for all descendants at a small levels $m$ can be evaluated explicitly, as an expansion in powers of $1/c$. The resulting expressions can be compared with the spectrum following from \eqref{result}, which we will write in terms of quantum numbers $n_k$ packaged as follows
\bea
 && \qquad   m_{p,r}\equiv \sum_k k^p n_k^r, \qquad m_p\equiv m_{p,1},\qquad m\equiv m_1,\qquad h=\tilde{\Delta}/\tilde{c},\\
\label{lambda3}
{\sf Q}_3&=&
\tilde \Delta ^2+\tilde\Delta  \left(6\, m_1-\frac{1}{4}\right) +\tilde c \left(4\, m_3+\frac{1}{60}\right)+\\&&   \left(m_3-{3\over 2} m_2-{1\over 4}m_1\right) -\frac{3 }{2}  m_{2,2} + 3 m_1^2 + \frac{3 }{2} h(2m_1-m_0-m_{0,2}) +{3\over 320}+{\mathcal O}(1/\tilde c),\nonumber
\eea
and similarly 
\bea
\label{lambda5}
{\sf Q}_5&=&\tilde{\Delta }^3+\left(15 m_1-\frac{5}{8}\right) \tilde{\Delta }^2+ \tilde{\Delta }\,  \tilde{c} \left(20 m_3+ \frac{1}{12}\right)+
\tilde{c}^2 \left(8 m_5-\frac{1}{63}\right)+ \, \,\,\quad\qquad\qquad\qquad \\ \nonumber
&& \tilde{\Delta } \left({5\over 12}(-5 m_1 -42 m_2+44 m_3)-{35\over 2}m_{2,2}+25 m_1^2+{15\over 2}h(2m_1-m_0-m_{0,2})+{23\over 192}\right)+\\
 \nonumber && \tilde{c} \left(
{1\over 12}(m_1-10m_3-120 m_4+64 m_5)-10 m_{4,2}+20 m_1 m_3 -{85\over 6048}
\right) +\mathcal O(\tilde c^0), \nonumber
\eea
and 
\bea
\label{lambda7}
\nonumber
{\sf Q}_7&=&\tilde{\Delta }^4 + \tilde{\Delta }^3 \left( 28 m_1 - \frac{7}{6}\right) + \tilde{\Delta }^2\, \tilde{c} \left( 56 m_3 + \frac{7}{30} \right) + \tilde{\Delta }\, \tilde{c}^2 \left( \frac{224}{5} m_5 - \frac{4}{45} \right) +\tilde{c}^3 \left( \frac{64}{5} m_7 + \frac{2}{75} \right) + \\  \nonumber
&& \tilde{\Delta }^2 \left( {7\over 6}\left(-7 m_1-66 m_2+76 m_3\right)-77 m_{2,2}+ 98 m_1^2  +21 h(2m_1 -m_0-m_{0,2})+{259\over 480}  \right) + \\  \nonumber
&& \tilde{\Delta }\, \tilde{c} \left( {7\over 75}\left(7m_1-70 m_3-960m_4+688 m_5\right) -{448\over 5} m_{4,2}+{784\over 5} m_1 m_3 - \frac{167}{1080} \right) + \\ \nonumber 
&& \tilde{c}^2 \left({2\over 225}(-10 m_1+21 m_3-210m_5-3780m_6+1704 m_7)-{168\over 5}m_{6,2} +{112\over 5}(2m_1 m_5 +m_3^2)+\frac{77}{2160}\right)   \\  &+&\mathcal O(\tilde c^1).
\eea
We checked, these expressions are in agreement with the computer algebra generated spectrum for $m\leq 12$, which serves as a non-trivial consistency check of \eqref{result}.

\section{Miscellaneous results }
\label{sec:applications}
Explicit expression for the spectrum of quantum $Q_{2n-1}$ in large  $c$ limit opens the opportunity to make progress in a number of adjacent directions. In this section we discuss several applications of our results. 

\subsection{Thermal expectation values of $Q_{2n-1}$}
\label{sec:thermalev}
Our first application is toward thermal exaction  value of $Q_{2n-1}$,  i.e.~averaged over the CFT Gibbs ensemble $\langle Q_{2n-1}\rangle_q\equiv {\rm Tr}(q^{L_0-c/24}Q_{2n-1})$. This question appears naturally, though in a more complicated form, to calculate the  averaged value of $Q_{2n-1}$ over the KdV Generalized Gibbs Ensemble (see section \ref{sec:GGE} below),  if one wants to match the GGE chemical potentials to describe equilibration endpoint of some initial state. The expectation value $\langle Q_{2n-1}\rangle_q$, which is essentially the one-point function of $T_{2n}$ \eqref{qKdV} on the torus, exhibits modular properties and can be represented as a covariant differential operator acting on the CFT torus partition function \cite{maloney2018thermal}. In fact, one can average $Q_{2n-1}$ over a particular Verma module, 
$\langle Q_{2n-1}\rangle_\Delta \equiv {\rm Tr}_\Delta(q^{L_0-c/24}Q_{2n-1})$, where sum goes over all Virasoro descendants of the primary state $|\Delta\rangle$. This sum too is a modular object and can be evaluated with help of the same differential operator 
\bea
\label{D1}
\langle Q_{2n-1}\rangle_\Delta&=&{\mathcal D}_n \chi_\Delta, \qquad \chi_\Delta\equiv {\rm Tr}_\Delta(q^{L_0-c/24})={q^{\tilde{\Delta}- \frac{1}{24}}/ \eta},\\
{\mathcal D}_n&=&D^n+\sum_{j=1}^{n-1} P_n^j (c, q) D^{n-j-1},\qquad D^n= D_{2(n-1)}\dots D_2D_0,
\label{D2}
\eea
and $D_r=q \partial_q -{r\over 12}E_2$ is Serre derivative. Each $P_n^j$ is a degree $j$ polynomial in $c$ with each coefficient being a modular form of weight $2j+2$, 
\bea
 P_n^j (c, q)=\sum_{k=1}^{j+1} P^{(k)}_{n,j} \tilde{c}^{j-k+1}  E_{2j+2}^{(n,k)}(q).
\eea
Here $P_{n,j}^{(k)}$ are numerical coefficients and $E_{2j+2}^{(n,k)}$ is some  modular form, which is a linear combination of $E_4^{a} E_6^b$ with $4a+6b=2j+2$ for non-negative integer $a,b$, normalized such that $E_{2j+2}^{(n,j)}=1+O(q)$. 
For $j=1,2,3,4,6$ there is a unique modular form of the weight $2(j+1)$ and therefore for these $j$, independently of $n$ and $k$,  $E_{2j+2}^{(n,k)}=E_{2j+2}$ where
\bea
E_{2n}= 1+{2\over \zeta(1-2n)}\sigma_{2n-1},\qquad \sigma_{p}= \sum_{k=1}^\infty {k^p q^k\over 1-q^k}.
\eea
For instance, in the simplest case of $Q_3$ the operator $\mathcal{D}_2$ is given by
\bea
\langle Q_{3}\rangle_\Delta&=&{\mathcal D}_2 \chi_\Delta = \left[D^2 + \frac{c}{1440} E_4 \right]\chi_\Delta.
\eea
In this case $P_{2,1}^{(1)}=1/60$ and  $P_{2,1}^{(2)}=1/1440$. Explicit expressions for $\mathcal{D}_n$ for $n\leq 7$  were found in \cite{maloney2018thermal}.
For higher $n$ the  modular form $E_{2i+2}^{(n,j)}$ and  coefficients $P_{n,j}^{(k)}$ are  not known.  

Strictly speaking (\ref{D1},\ref{D2}) is an unproven ansatz proposed in \cite{maloney2018thermal}. We find it to be consistent with the  large $c$ spectrum of $Q_{2n-1}$ \eqref{result} and fix  two leading  in $c$ terms in $P_n^j$. To compare with \eqref{D1}, we need to calculate $\langle Q_{2n-1}\rangle_\Delta$ starting from \eqref{result}. Here the following straightforward  identities will be helpful 
\bea
\label{sigmaid1}
&&\langle  \sum_{k=1}^\infty n_k k^p \rangle_\Delta =\sigma_p \chi_\Delta,\quad \langle  \sum_{k=1}^\infty n^2_k k^p \rangle_\Delta = \left( 2q \partial_q \sigma_{p-1}-\sigma_p \right)\chi_\Delta,\quad  \\ &&\langle  \sum_{k=1}^\infty n_k k^p \sum_{\ell=1}^\infty n_\ell \ell^{p'} \rangle_\Delta = \left(q \partial_q \sigma_{p + p'-1} + \sigma_p \sigma_{p'} \right) \chi_\Delta, \label{sigmaid2}
\eea
where by $n_k$ we mean the  quantum  numbers \eqref{Qeig}. Then \eqref{leadingQ} immediately yields 
\bea
\langle Q_{2n-1}\rangle_\Delta = \tilde{\Delta}^n \chi_\Delta+\sum_{j=0}^{n-1} \tilde{\Delta}^{n-p-1} \tilde{c}^p \xi_n^p \left( \sigma_{2p+1} +{\zeta(-2p-1)\over 2} \right)\chi_\Delta+O(\tilde{c}^{n-2}), \label{expansionTrQ}
\eea
where we assumed the usual limit,  $h=\tilde{\Delta}/\tilde{c}$ is kept fixed while $\tilde{c}\rightarrow \infty$. Comparing this with \eqref{D1}, we immediately see that the leading $\tilde{\Delta}^n$ term is coming from  (we drop $\chi_\Delta$ for simplicity)
\bea
D^n  \rightarrow (q \partial_q)^n \rightarrow \tilde{\Delta}^n . 
\eea
Similarly we can trace origin of all $\tilde{c}^{n-1}$ terms, 
\bea
\nonumber
D^n\rightarrow (q \partial_q)^{n}- {n(n-1)\over 12}E_2 (q \partial_q)^{n-1} \rightarrow  \tilde{\Delta}^{n-1} n\left(
 \sigma_{1} -{1\over 24}\right)  -\tilde{\Delta}^{n-1}{n(n-1)\over 12}E_2=-{n(2n-1)\over 24}E_2, \nonumber
\eea
which agrees with \eqref{expansionTrQ},
and 
\bea
P^{(1)}_{n,j} \tilde{c}^j E^{(n,1)}_{2j+2} D^{n-j-1} \rightarrow  P^{(1)}_{n,j} \tilde{c}^j E^{(n,1)}_{2j+2} (q\partial_q )^{n-j-1} \rightarrow 
P^{(1)}_{n,j} \tilde{\Delta}^{n-j-1} \tilde{c}^j E^{(n,1)}_{2j+2} ,
\eea
for $n-1\geq j>0$. From here immediately follows 
\bea
\label{E1}
P^{(1)}_{n,j}=\tilde{R}^{(1)}_{n,j},\qquad  E_{2j+2}^{(n,1)}=  E_{2j+2},\qquad n-1\geq j\geq 1.
\eea

To fix $P^{(2)}_{n,j}$ it is convenient to take $q\rightarrow 0$ limit and compare $\langle Q_{2n-1}\rangle_\Delta$ with \eqref{vacuumE}, yielding
\bea
\nonumber
&&P^{(2)}_{n,1}=\tilde{R}^{(2)}_{n,0}-\frac{n (n-1) (12 n^2-16 n-1)}{3456}={n(n-1)(12 n^2-38 n+31)\over 8640},\\
\nonumber
&&P^{(2)}_{n,j}=\tilde{R}^{(2)}_{n,j-1}+{(n-j)(2(n-j)-1)\over 24} P^{(1)}_{n,j-1},\quad n-1\geq j\geq 2.
\eea
Evaluation of $E^{(n,2)}_{2j+2}$ is a more challanging  task and requires first using (\ref{sigmaid1},\ref{sigmaid2}) and then combining pieces into modular forms to match (\ref{D1},\ref{D2}). We note,  there are terms in \eqref{result} proportional to $\tilde{\Delta}^{n-1}\tilde{c}^{-1}$, but \eqref{D1} has no negative powers of $c$.
Hence these terms must vanish after averaging, which  follows from the identity $q\partial_q \sigma_{-1}-\sigma_1=0$ and serves as a consistency check. 
The final expression reads
\bea
\label{E2}
&&P^{(2)}_{n,j} E^{(n,2)}_{2j+2}= {(2n-1)\sqrt{\pi} \Gamma(n+1)\over 8\Gamma(j+3/2)\Gamma(n-j)}
\left( ((2n-1) y_1(j)-2j-1){\zeta(-2j-1)\over 2}E_{2j+2}\, \, - \right. \\   &&(n-1-j)\zeta(-2j+1) D_{2j}E_{2j}
+ \left.   {(2n-1)\over 4}\sum_{s=1}^{j-2} \zeta(-2s-1)\zeta(-2(j-s)+1) E_{2s+2}E_{2(j-s)}
\right). \nonumber
\eea
It is valid  for $n-1\geq j\geq 2$. For $j=1$, there is a unique modular form $E^{(n,2)}_{2j+2}=E_4$.
Also, as was mentioned above  $E^{(n,2)}_{2j+2}=E_{2j+2}$ for $j=2,3,4,6$, which can be checked straightforwardly. 
Because of the identities between modular forms there are other ways to write \eqref{E2}.

Explicit form of ${\sf Q}_{2n-1}^0$ up to $\tilde{c}^{n-3}$ order allows us, in principle, to calculate $P^{(3)}_{n,j}$, although calculation of $E^{(n,3)}_{2j+2}$ would require first extending \eqref{result} to the next $1/c$ order. Given involved form of $P^{(2)}_{n,j}$ and  $E^{(2)}_{n,j}$ we do not expect the answer to be simple.

\subsection{Generalized Gibbs Ensemble}
\label{sec:GGE}
Spectrum of $Q_{2n-1}$ can help understand the qKdV generalized Gibbs ensemble (GGE)
\bea
\rho_{\rm GGE}=e^{-\sum_n \mu_{2n-1}\, Q_{2n-1}},\qquad Z_{\rm GGE}=\Tr \rho_{\rm GGE},
\eea 
and corresponding  (generalized) partition function and free energy. Earlier attempts to evaluate  KdV generalized free energy include \cite{de2016remarks,maloney2018generalized,GGE, GGE2}. 
The GGE describes local equilibrium  in a state carrying specific values of qKdV charges. It is expected  on general grounds that most initial states, upon equilibration, can be locally described by the GEE with the appropriate values of chemical potentials $\mu_{2n-1}$ \cite{rigol2007relaxation}.
From the mathematical point of view, it is of great interest to investigate modular properties of $Z_{\rm GGE}$,  generalizing modular invariance of the conventional torus partition function $\mu_{2n-1}=0$, for $n>1$. 

The explicit spectrum of $Q_{2n-1}$ in the large $c$ limit allows in principle  to calculate the generalized sum over a particular Verma module
\bea
\Tr_{\Delta} e^{-\sum_n \mu_{2n-1} Q_{2n-1}}
\eea
in the ``holographic limit'': $h=\tilde{\Delta}/\tilde{c}$, $t_n:=\mu_{2n-1}/\tilde{c}^{n-1}$ fixed, $\tilde{c}\rightarrow \infty$, by expanding the answer in powers of $1/c$.  In practice sums of exponents of quadratic or higher order expressions in $n_k$ 
\bea
\sum_{n_k} e^{O(n^2)} 
\eea
can not be evaluated, and we restrict our analysis to first non-trivial $1/c$ order, 
\bea
\nonumber
\Tr_{\Delta} e^{-\sum_n \mu_{2n-1} Q_{2n-1}}=e^{-\tilde{c}\sum_n t_n h^n} {e^{-\sum_n t_n \sum_{p=0}^{n-1} h^{n-1-p}\xi_n^ p \zeta(-2p-1)/2}\over \prod_{k=1}^\infty \left(1-e^{-k \sum_n (2n-1)n\, t_n\,   h^{n-1} {}_2 F_1(1,1-n,3/2,-k^2/h) }\right)}.
\eea
From here generalized  partition function can be evaluated using Cardy formula (we are only writing explicitly the chiral part), 
\bea
\label{ZGGE}
Z_{\rm GGE}&=&e^{\tilde{c}f_0+ f_1+O(1/\tilde{c})}, \\  \label{f0}
f_0&=&\sum_{n=1}^\infty (2n-1)\, t_n\, h^n,\\ 
h^{1/2} &=&{1\over 2\pi} \sum_{n=1}^\infty t_{n}\, n\, h^n, \label{heq} \\
\label{f2}
f_1&=&  -\sum_{k=1}^\infty \ln \left(1-e^{-\gamma}\right)-\sum_{n=2}^\infty t_{n}\, h^{n-1}\left( \sum_{p=0}^{n-1} \xi_n^p\, h^{-p} {\zeta(-2p-1)\over 2}- {n\over 24} \right),\\
\gamma(k)&=&k \sum_{n=1}^\infty (2n-1)n\, t_n\,   h^{n-1} {}_2 F_1(1,1-n,3/2,-k^2/h).  \label{Zend}
\eea
Here  $Z_{\rm GGE}$ is understood to be a function of $t_n\equiv \mu_{2n-1} \tilde{c}^{1-n}$, while $h$ is a function of $t_n$ satisfying \eqref{heq}. For \eqref{ZGGE}-\eqref{Zend} to be valid, resulting $\tilde{\Delta}= \tilde{c}\, h$ should be in the regime of validity of Cardy formula. There are at least two limits when this assumption is controllable. First,  \eqref{ZGGE} is valid for any large $c$ theory in the thermodynamic limit.  We introduce the spatial circle radius $L$ (we kept $L=1$ in the paper so far) and inverse temperature $\beta$, $\mu_1=t_1=\beta/L$. By taking $L \rightarrow \infty$, while all other chemical potentials scale as $\mu_{2n-1} \propto  t_n \sim L^{1-2n}$ to ensure that values of all $Q_{2n-1}\propto L$ are extensive, we find the saddle point value $h \sim L^2$ and $f_0, f_1 \sim \L$. (The scaling of $f_1$ follows  by substituting the  sum over $k$ in \eqref{f2} by an integral over $\kappa=k^2/h$.) In this limit second term in \eqref{f2}, the sum over $n$,  is  sub-extensive and can be neglected. We therefore arrive at the leading (extensive) contribution to $f_0$ and $f_1$ found in \cite{GGE2}.

Second case when  \eqref{ZGGE}-\eqref{Zend} can be trusted is in  holographic theories, i.e.~large $c$ theories satisfying HKS sparseness condition \cite{hartman2014universal}. From the holographic point of view $f_0$ is the free energy of  BTZ black hole in the Euclidean classical theory of gravity with the deformed boundary conditions such that the dual CFT Hamiltonian is $H=\sum_n \mu_{2n-1} Q_{2n-1}$ \cite{perez2016boundary,GETH,Dymarsky_2020}. The leading correction $f_1$ can be interpreted as the one-loop contribution coming from the boundary gravitons. Different solutions of \eqref{heq} means Euclidean path integral could have numerous BTZ saddles and the  condition $h(t_n)>1/12$ necessary for the validity of Cardy formula would come automatically as the requirement of smoothness of  bulk geometry.

It is possible to fine-tune chemical potentials $t_n$ such that $\gamma(k)$ \eqref{Zend} for some $k$ will vanish. That will render $f_1$ divergent, indicating higher order $1/c$ corrects are necessary to make free energy finite. Schematically, the spectrum ${\sf Q}_{2n-1}$ is an expansion in $n_k/\tilde{c}$. For the higher order corrections to contribute at the leading order, the quantum numbers $n_k$ should be of order $\tilde{c}$. In terms of  the classical problem of section \eqref{sec:classic},  action variables $I_k$ should be of order one rather than infinitesimal. In other words, leading contribution would come from a non-trivial saddle when classical $u(\varphi)$ is not a constant but some solitonic solution. Such saddles, describing  black holes, which are geometrically different from the BTZ configurations, were constructed in \cite{Dymarsky_2020} and it was  shown that  for certain parameters $\mu_{2n-1}$ they give leading contribution to generalized free energy. We dubbed these configurations ``KdV-charged'' black holes to  emphasize that higher KdV charges $Q_{2n-1}$, even at leading order in $c$, are different from $Q_1^n$, unlike for BTZ configurations for which  $u(\varphi)=u_0$ is a constant and  $Q_{2n-1}\sim u_0^n$.

Theoretical control over generalized free energy  in the large $c$ limit can be used to probe modular properties of $Z_{\rm GGE}$. The currents $T_{2n}$ \eqref{qKdV} have no anomalous dimension and therefore naively $Z_{\rm GGE}$ should be invariant under modular transformation $t_1\rightarrow t_1'=(2\pi)^2/t_1$ accompanied by 
\bea
\label{transform}
t_n \rightarrow (-1)^n \left({2\pi\over t_1}\right)^{2n} t_n,\quad n>1.
\eea
This only holds to linear order in $t_n, n>1$, i.e.~at the level of thermal expectation values $\langle Q_{2n-1}\rangle_q$ discussed in section \ref{sec:thermalev}. At higher orders invariance is broken due to colliding $T_{2n}$ \cite{maloney2018thermal}. To restore invariance of $Z_{\rm GGE}$, while working in the $c\rightarrow \infty$ limit  one may require $f_0$ given by \eqref{f0},\eqref{heq} to be invariant under the hypothetical transformation  $t_n\rightarrow t_n'(t_n,t_1)$, $n>1$. More accurately, in addition to BTZ black holes described by \eqref{f0},\eqref{heq} we should include vacuum (thermal AdS${}_3$) and KdV-charged black holes to the list of possible saddles. Given a non-trivial diagram of the Hawking-Page phase transitions, to match leading saddles, the hypothetical transformation   $t_n\rightarrow t_n'(t_n,t_1)$ should be very complicated, with numerous branches of continuity. This may indicate  that in the presence of higher KdV charges modular invariance of $Z_{\rm GGE}$ is not mathematically natural. Similar conclusion is recently reached in \cite{Downing:2021mfw}, which evaluated $Z_{\rm GGE}$ explicitly in the case of $c=1/2$ free fermion model. They found that to reproduce $Z_{\rm GGE}$ in the dual channel, one needs to sum over not one but three fermion Hilbert spaces, schematically $Z_{\rm GGE}(t_1,t_3) \propto Z_1(t_1') Z_2(t_1') Z_3(t_1')$, a mathematical observation (conjecture), which so far has no  physical interpretation. To summarize, failure to establish invariance of  $Z_{\rm GGE}(t)$ under modular transformation supplemented by an appropriate map $t_n \rightarrow t_n'$ in both infinite $c$  limit and for $c=1/2$ model may suggest that it is not  mathematically natural and instead covariance of $Z_{\rm GGE}(t)$ under \eqref{transform} should be investigated. 

\subsection{Transfer Matrix}
\label{sec:TM}
In the classical case, as follows from \eqref{kdveq}, charges $Q_{2n-1}$ encode  asymptotic expansion of the quasi-momentum $p(\lambda)$. The quasi-momentum controls the eigenvalues $e^{ \pm 2\pi i p(\lambda)}$ of the monodromy matrix of the differential equation \eqref{sch}. Instead of $p(\lambda)$  one can consider the trace of monodromy matrix 
\bea
T(\lambda)=2 \cos(2\pi p(\lambda)). \label{classicalTM}
\eea
In case of the constant potential $u(\varphi)=h$ this becomes $T(\lambda)=2 \cos(2\pi\sqrt{\lambda-h/4})$.

In quantum case $T(\lambda)$ becomes the transfer matrix, which is related to qKdV charges via an asymptotic expansion \cite{bazhanov1996integrable} 
\bea
\label{lnTdef}
\ln T &=& \kappa\, \mu^{1/2}\, {\rm 1} -\sum_{n} C_n\, \mu^{1/2-n}\, Q_{2n-1},\quad \mu\rightarrow \infty,
\eea
where 
\bea
    \kappa&=&\frac{2\sqrt{\pi}\Gamma\left(\frac{1}{2}-\frac{\xi}{2}\right)}{\Gamma\left(1-\frac{\xi}{2}\right)}
    \left(\Gamma\left(1-\beta^2\right)\right)^{1+\xi},\\     \label{coef:C_n}
    C_n&=&\frac{\sqrt{\pi}(1+\xi)\beta^{2n}\Gamma\left((n-\frac{1}{2})(1+\xi)\right)}{\Gamma(n+1)\Gamma\left(1+(n-\frac{1}{2})\xi\right)}
    \left(\Gamma\left(1-\beta^2\right)\right)^{-(2n-1)(1+\xi)},\\
 \beta&\equiv& \sqrt{\frac{1-c}{24}}-\sqrt{\frac{25-c}{24}},
    \qquad 
    \xi\equiv \frac{\beta^2}{1-\beta^2}.
\eea
Variable $\mu$ will become spectral parameter $-\lambda$ in the classical limit. The original paper \cite{bazhanov1996integrable} introduces another variable $\lambda$ , defined as $\mu\equiv \lambda^{2(1+\xi)}$. We use this definition  in the reminder of this section. 

We are interested in the limit $c\rightarrow \infty$, or $\beta \rightarrow 0$. Following \cite{bazhanov1996integrable} we introduce $p^2=\beta^2 \tilde{\Delta}$ which remains finite in this limit finite, $p^2 \rightarrow -\tilde{\Delta}/4\tilde{c}=-h/4$.  (This is, obviously, a different quantity from the quasi-momentum $p(\lambda)$ mentioned above.) 
We would like to find $T$ by summing the asymptotic expansion \eqref{lnTdef} while expanding it in powers of $\beta^2$ which corresponds to $1/c$ expansion. In principle we can use the spectrum \eqref{result} to calculate $\ln T$ acting on an excited state, but resort to a simpler calculation for $\ln T$ acting on a primary state. In this case $Q_{2n-1}$ in \eqref{lnTdef} should be substituted by ${\sf Q}_{2n-1}^0$, which we expand in powers of $\beta^2 \propto 1/c$ \eqref{vacuumE}.  The calculation is tedious  and we only give the final expression 
\bea
\label{lnTanswerr}
(\ln T)_{\rm asympt}|\Delta\rangle &=&2\pi i \sqrt{p^2-\lambda^2} \Phi(\lambda,p)|\Delta\rangle,\\
\nonumber
 \Phi(\lambda,p)&=&1+\beta^2\Psi-\frac{\beta^4}{48 \lambda^2(p^2-\lambda^2)^3}\left[3\lambda^4 p^2+2\pi^2 \lambda^4(p^2-\lambda^2)(4p^2-3\lambda^2)\right]
    \\
    &&\quad +\frac{\beta^4}{2 \lambda^2}\left[
    (2p^2+\lambda^2)\Psi^2-2(p^2-\lambda^2)\lambda\Psi \frac{d\Psi}{d \lambda}
    \right]+\mathcal{O}(\beta^6),\\
 \Psi(\lambda,p)
   &=&\frac{\lambda^2}{\lambda^2-p^2}\left[\gamma
    +\frac{1}{2}\psi\left(2\sqrt{p^2-\lambda^2}\right)
    +\frac{1}{2}\psi\left(-2\sqrt{p^2-\lambda^2}\right)
   \right]
   .
\eea
Here $\psi$ is the polylog function. 

Given analytic form of \eqref{lnTanswerr} it is tempting to extend its validity   from the  asymptotic regime $\lambda \rightarrow \infty$ to the vicinity of $\lambda=0$. This is  clearly wrong as even in the strict classical limit $\beta\rightarrow 0$ we do not recover correct  classical expression for the trace of monodromy matrix simply from $e^{(\ln T)_{\rm asympt} }$. Yet  in the limit $\beta\rightarrow 0$ the correct answer is reproduced by the following simple conjectural expression (we implicitly assume this is an eigenvalue of $T$ acting on $|\Delta\rangle$),
\bea
T_{\rm guess}(\lambda, \beta, p)=e^{(\ln T)_{\rm asympt} }+e^{-(\ln T)_{\rm asympt} }, \label{T}
\eea 
and we would like to check if it  could be valid beyond the strict $\beta=0$ limit. 
To that end we expand \eqref{T} in powers of $\beta$ (amended by an expansion in $\lambda$), to find 
\bea
\nonumber
T_{\rm guess}=2\cos(2\pi \sqrt{p^2-\lambda^2})+\beta^2\left[
\frac{2\pi\sin(2\pi p)}{p}\left(2\gamma+\psi(2p)+\psi(-2p)\right)\lambda^2+  \right. \\ \left. \nonumber
\left(-\frac{2\pi^2 \cos(2\pi p)}{p^2}\left[2\gamma+\psi(2p)+\psi(-2p)\right) \right. \right. \\ \left. \left. \nonumber
 + \frac{\pi \sin(2\pi p)}{p^3}\left(2\gamma+\psi(2p)+\psi(-2p)-2p\psi^{(1)}(2p)+2p\psi^{(1)}(-2p)\right) \right)\lambda^4+
 O(\lambda^6)\right]+\\
\beta^4 \left[\frac{\pi \sin(2\pi p)}{12 p^3}\left[3+8 \pi^2 p^2+12p^2\left(2\gamma+\psi(2p)+\psi(-2p)\right)^2\right]\lambda^2 +O(\lambda^4)\right]+O(\beta^6). \nonumber
\eea
Small $\lambda$ expansion of the actual $T$ is given in \cite{bazhanov1996integrable} in the explicit form in terms of the integrals of free field correlators. A comparison with $T_{\rm guess}$ reveals that, besides the classical $\beta^0$  term, which matches the classical expression \eqref{classicalTM} for a constant potential  $u=h$, only $\beta^2\lambda^2$ term coincides with, while $\beta^2 \lambda^4$ and $\beta^4 \lambda^2$ terms do not match the correct result. We thus conclude that the conjectural expression \eqref{T} is missing non-perturbative terms, which are not captured  by the asymptotic expansion \eqref{lnTdef}.

\section{Discussion}
\label{sec: discussion}
In this paper we obtained spectrum of quantum KdV charges $Q_{2n-1}$ in first two non-trivial orders in $1/c$ expansion.  Our result \eqref{resultpreliminary}  and \eqref{result} is valid  in the semiclassical   limit of large central charge $c\rightarrow \infty$ with the ratio of  $\Delta/c$ kept fixed.  This limit is inspired by holographic correspondence, when CFT is dual to weakly coupled gravity. Accordingly, dynamics of stress-energy sector becomes semiclassical, with the leading (classical) contribution governed by integrable dynamics on the co-adjoint orbit of the Virasoro algebra. 
Under semiclassical quantization  classical action variables $I_k$ are promoted to  integer quantum numbers $n_k$, and the spectrum of $Q_{2n-1}$ looks most elegant in terms of variables $\tilde{\Delta}$ and $\tilde{c}$ \eqref{quantization}. At each order in $1/\tilde{c}$ the quantum answer is a polynomial in $n_k$. Classical calculation fixes the leading  term with the highest power of $n_k$, while all other terms should be regarded as ``quantum corrections.'' 
We have seen that semiclassical quantization, combined with the values of qKdV charges $Q_{2n-1}$ acting on primary states, is sufficient to completely fix these quantum corrections and obtain the spectrum of excited states at least in first two orders in $1/c$. We conjecture  this quantization scheme can be extended to higher orders in $1/c$.
We laid the groundwork for the next order $1/c^3$ by calculating classical $Q_{2n-1}(h,I_k)$ as well as ``energies'' on primary states ${\sf Q}^0_{2n-1}$, albeit in the latter case not all terms are known analytically.  To complete the job one would need to find analytic expressions for ${\sf Q}^0_{2n-1}$ and develop a dictionary that maps each term to an infinite sum, yielding this term  back via zeta-function regularization. 

It is tempting to interpret quantization of $Q_{2n-1}$ holographically, as a semiclassical quantization of boundary gravitons in AdS${}_3$. We develop this picture at first $1/c$ order in the appendix \ref{sec:holography}, but  holographic picture does not provide any immediate insight into ``quantum corrections'' appearing at  higher orders  in $1/c$.

The obtained spectrum has several immediate applications. First,  in section \ref{sec:thermalev} we calculated two leading terms in large $c$ expansion of  the ``thermal expectation values'' $\langle Q_{2n-1}\rangle_\Delta \equiv {\rm Tr}_\Delta(q^{L_0-c/24}Q_{2n-1})$, where sum goes over a particular Verma module, and compared them with the predictions of \cite{maloney2018thermal}. Covariance under modular transformation of $\langle Q_{2n-1}\rangle_\Delta$ in each order in $1/c$ serves as a non-trivial check of our main result \eqref{result}. We also fixed two leading terms in the differential operator ${\mathcal D}_n$ yielding thermal expectation values via  $\langle Q_{2n-1}\rangle_\Delta={\mathcal D}_n {\rm Tr}_\Delta(q^{L_0-c/24})$, see \eqref{E1} and \eqref{E2}. Second, in section \ref{sec:GGE} we calculated first $1/c$ correction to generalized free energy of the  qKdV Generalized Gibbs Ensemble
\bea
Z_{\rm GGE}=\Tr e^{-\sum_n \mu_{2n-1} Q_{2n-1}}.
\eea 
The latter describes local equilibrium of a 2d CFT in a state carrying specific values of qKdV charges. It is of great interest to further investigate mathematical properties of $Z_{\rm GGE}$, in particular covariance under modular transformation. Third, in section \ref{sec:TM} using asymptotic expansion we calculated quantum transfer matrix acting on a primary state in first two non-trivial orders in  $1/c$ expansion. Unfortunately the obtained expression is lacking terms non-perturbative in spectral parameter, which can not be fixed from the knowledge of spectrum of $Q_{2n-1}$ alone. 

There are several potential applications of our results, which we hope to address in the future. The obtained spectrum of $Q_{2n-1}$ will be helpful to study generalized Eigenstate Thermalization Hypothesis  of 2d CFTs \cite{GETH} at the subleading order in $1/c$. We also expect the semiclassical quantization approach developed in this paper could  be helpful in the context of Intermediate Long Wave hierachry, which is closely related to qKdV problem. 
More generally, it would be interesting to bridge the gap between the semiclassical approach of this work with the Bethe anzatz  approach of \cite{LitvinovEq} by taking ``holographic'' limit $c\rightarrow \infty$ with fixed $h=\tilde{\Delta}/\tilde{c}$ of the appropriate Bethe anzatz equations.

\acknowledgments
We thank A.~Gorsky, I.~Krichiver, A.~Litvinov, N.~Nekrasov,  A.~Okounkov, V.~Pestun and A.~Zamolodchikov for helpful discussions.
AD,  AK, and SS  were supported by the National Science Foundation under Grants No.~PHY-1720374 and PHY-2013812. AD is grateful to IHES and KITP for hospitality, where this work was partially done.  The research at KITP was supported in part by the National Science Foundation under Grant No.~PHY-1748958.  The research at IHES  was funded from the European Research Council (ERC) under the European Union's Horizon 2020 research and innovation program (QUASIFT grant agreement 677368). KP was supported by the RFBR grant  19-32-90173. 

\appendix
\section{Spectrum of linear perturbations from AdS${}_3$} \label{sec:holography}
In the classical (infinite central charge) limit gravity in AdS${}_3$ can be described in terms of two functions $u(t,\varphi)$ and $\bar{u}(t,\varphi)$ living at the boundary and satisfying EOM, $\dot{u}=\partial_\varphi u$ and  $\dot{\bar{u}}=-\partial_\varphi \bar{u}$. That is in the conventional case, when the dual CFT's Hamitlonian is $H=Q_1+\bar{Q}_1=L_0+\bar{L}_0-{c/12}$. Should the Hamiltonian be chosen to be one of the higher qKdV charges, $H=Q_{2n-1}+\bar{Q}_{2n-1}$, functions $u,\bar{u}$ will be satisfying higher KdV equations 
\bea\dot{u}={c\over 24}\{Q_{2n-1},\delta u\}=(2n-1) \partial R_n,\label{eom}
\eea and similarly for $\bar u$ \cite{perez2016boundary,Ojeda:2019xih,Dymarsky_2020}. In this case the spectrum of $Q_{2n-1}$ is the spectum of small fluctuations of $u$ above the constant backgroun $u=u_0$, that corresponds to unpertubed metric in AdS${}_3$. In other words,  to quantize $Q_{2n-1}$ we consider linearized EOM for small fluctuations  $u=u_0+\delta u$, where $\delta u\propto e^{i\varepsilon t+i k\varphi}$ is a flat wave. We want to find energy $\varepsilon$ of the flat wave which satisfies the equation of motion \eqref{eom}
\begin{equation}
    i\varepsilon_n \delta u =(2n-1)\delta \partial R_n.
\end{equation}
For example in the case $n=1$ we have $\varepsilon_1=k$, in case $n=2$ we have $\varepsilon_2=2u_0 n+{4\over 3}k^3$ and so on. In general we can get from \eqref{iterativeR}
\begin{eqnarray}
\delta \partial R_{n+1}={n+1\over 2n+1} \delta (\partial u+2 u\partial -2\partial^3) R_n=i{n+1\over 2n+1}(k R_n(u_0)+2(u_0+k^2)\delta \partial R_n) .\quad 
\end{eqnarray}
Hence we find the following iterative relation for $\varepsilon_n$
\begin{equation}
\varepsilon_{n+1}=(2n-1){n+1\over 2n+1}\left[2(u_0+k^2)\varepsilon_n+k u_0^n\right],   
\end{equation}
where we have used that $R_n(u_0)=u_0^n$.
Each $\varepsilon_n$ is a polynomial of the form 
\begin{equation}
    \varepsilon_n=\sum_{p=0}^{n-1} \xi^p_n\, k^{2p+1} u_0^{n-1-p}, 
\end{equation}
where $\zeta_n^p$ satisfy
\begin{equation}
    \xi_{n+1}^p=(2n-1){2(n+1)\over 2n+1} \left(\xi_n^p+\xi_n^{p-1}\right),
\end{equation}
and we defined $\xi_n^{-1}\equiv 1/2$. The solution  is easy to find, cf.~\eqref{xidef},
\begin{equation}
    \xi_n^p={(2n-1)\Gamma(n+1)\Gamma(1/2)\over 2\Gamma(p+3/2)\Gamma(n-p)}.
\end{equation}

To match the spectrum of individual bosons $\varepsilon_n(k,u_0)$ with the spectrum of quantum $Q_{2n-1}$ we need to restore powers of $\tilde{c}$ and make the following identification
\begin{equation}
\label{spectrumH}
    Q_{2n-1}=\tilde{\Delta}^n+ \tilde{c}^{n-1} \sum_k \left(n_k+{1\over 2}\right) \varepsilon_n(k,u_0)+\dots
\end{equation}
where $n_k$ are boson occupation numbers of boundary gravitons and $u_0=\tilde{\Delta}/\tilde{c}$. This  reproduces the spectrum of $Q_{2n-1}$ at two first leading orders in $1/c$ and provides physical interpretation of $n_k$. Unfortunately the holographic picture  provides no clear path to compute higher $1/c$ corrections to \eqref{spectrumH}.

\section{Brute-force pertubative calculation}
\label{sec:pert}
A straightforward but a laborious approach to evaluate $Q_{2n-1}$ in terms of action variables $I_k$ would be to use Fourier modes  $u_k$ of $u$, 
\begin{equation}
    u(\varphi)=\sum_k u_k e^{i k\varphi},
\end{equation}
to parametrize the co-adjoint orbit of Virasoro algebra, i.e.~the space of potentials $u$ sharing the same orbit invariant $h$ \eqref{hdef}. To that end $u_0$ should be understood as a function of $u_k$ \cite{Witten}. Then $Q_{2n-1}$ and $I_k$  can be expressed in terms of $u_k$, and consequently in terms of each other. 

In terms of the Fourier modes the Poisson bracket is 
\begin{equation}
   i\frac{c}{24} \{u_k,u_\ell\}=(k-\ell)u_{k+\ell}+ 2k^3 \delta_{k+\ell}.
\end{equation}
This coincides with the Virasoro algebra upon $u_0$ is shifted by a constant. At this point we introduce the orbit invariant $h(u_k)$ and express it  in terms of $u_k$ by expanding in power series
\begin{equation}
   h=u_0+\sum_{n=2}^\infty U_{n},\qquad U_{n}=
    \frac{1}{n!}\sum_{\substack{p_1,\cdots, p_n\\ o_1+\dots+p_n=0,\, p_k\neq 0}}h_{p_1,\cdots,p_n}u_{p_1}\cdots u_{p_n}.
\end{equation}
After imposing  ${c\over 24}\{h(u),u_k\}=0$ for any $k$ we find 
\bea
\nonumber
h_{p_1,p_2}&=&.-\frac{p_1^2+p_2^2+2h}{4 (p_1^2+h)(p_2^2+h)},
    \\  \nonumber
    h_{p_1,p_2,p_3}&=&\frac{p_1^2+p_2^2+p_3^2+6h}{8 (p_1^2+h)(p_2^2+h)(p_3^2+h)},\\ \nonumber
h_{p_1,p_2,p_3,p_4}&=&-
   \frac{15 h^4- 25 h^3 q_2 +h^2 (13 q_2^2-9 q_4)+h(-3(q_2^3+q_3^2)+8 q_2 q_4)+q_2^2 q_4-q_2 q_3^2-4 q_4^2}{den},\quad 
\eea
where 
\bea
&&q_2\equiv p_1 p_2 + p_1 p_3 + p_2 p_3 + p_1 p_4 + p_2 p_4 + p_3 p_4=-\frac{1}{2}\sum p_k^2,
\\
&&q_3 \equiv p_1 p_2 p_3 + p_1 p_2 p_4 + p_1 p_3 p_4 + p_2 p_3 p_4=\frac{1}{3}\sum p_k^3,
\\
&&q_4 \equiv p_1 p_2 p_3 p_4 =-\frac{1}{4}\sum p_k^4+ \frac{1}{2} q_2^2,\\ \nonumber
&&den=
   [(p_1+p_2)^2+(p_3+p_4)^2+2h][(p_1+p_3)^2+(p_2+p_4)^2+2h][(p_1+p_4)^2+(p_2+p_3)^2+2h]\times \\ &&\qquad \prod_{k=1}^4 ((p_k^2+h). 
\eea
Now we can get rid of $u_0= h- \sum_{n=2}^\infty U_n$  and express the Poisson brackets in terms of $u_k$, $k\neq 0$, 
\begin{equation}
  i \frac{c}{24} \{u_k,u_\ell\}=\delta_{k+\ell}(2k)(k^2+h-\sum_{n=2}^\infty U_n)+(1-\delta_{k+\ell})(k-\ell) u_{k+\ell}.
   \label{Poisson_full}
\end{equation}
Our next goal is to find symplectic form associated with the Poisson brackets
\begin{align}
\omega =\frac{c}{24} \times  \frac{i}{2}  \sum_{k\neq 0, \ell\neq 0} \omega_{k,\ell}\, du_k \wedge du_\ell,\qquad 
\omega_{k,\ell}= \sum_{n=0}^\infty \omega^{(n)}_{k,\ell},
\end{align}
where $\omega^{(n)}_{k,\ell}$ is an order $n$ homogeneous polynomial in $u_k$.  We find, order by order in $u_k$, 
\bea
    \omega^{(0)}_{k,\ell}&=& -\frac{1}{2k(k^2+h)}\delta_{k+\ell},\\
\omega^{(1)}_{k,\ell}
&=&
-(1-\delta_{k+\ell}) \frac{k-\ell}{4k\ell (k^2+h)(\ell^2+h)} u_{-k-\ell},\\
\omega^{(2)}_{k,\ell}&=&-
 \frac{1}{8k \ell (k^2+h)(\ell^2+h)} \sum_{m\neq 0}
 \left[\frac{k \delta_{k+\ell}}{m^2+h}u_m u_{-m}
 +\right. \\ 
 &&  \quad \left. (1-\delta_{k-m})(1-\delta_{\ell+m})\frac{(k+m)(\ell-m)}{m(m^2+h)}u_{-k+m}u_{-\ell-m}
 \right]
\eea
We now would like to introduce (rescaled) normal coordinates $z_k$ near the origin $u_k=0$ (which corresponds to constant $u(\varphi)=h$), such that 
\begin{align}
\frac{24}{c}\omega &=  \frac{i}{2}\sum_{k\neq 0}
\frac{-1}{2k(k^2+h)}   d{z}_k \wedge d {z}_{-k},\qquad i \frac{c}{24} \{{z}_k,{z}_\ell\}=\delta_{k+\ell}(2k)(k^2+h).
\end{align}
We find 
\begin{align}
{z}_k 
 &= u_k
 +\frac{1}{4}\sum_{\substack{p_1+p_2=k\\ p_i\neq 0}}\frac{1}{p_1 p_2}u_{p_1}u_{p_2}
 +\frac{1}{24}\sum_{\substack{p_1+p_2+p_3=k\\p_i\neq 0,k}}\frac{p_1p_2+p_2p_3+p_3p_1}{ p_1 p_2 p_3(p_1+p_2)(p_2+p_3)(p_3+p_1)}u_{p_1}u_{p_2}u_{p_3} \nonumber
 \\
 &\quad
 -\frac18\sum_{\ell\neq 0,\pm k}\frac{2\ell^2+h}{ \ell^2(k^2-\ell^2)(\ell^2+h)}u_ku_{\ell}u_{-\ell}
 -\frac{2k^4-k^2 h +h^2}{32k^4(k^2+h)^2}u_k^2 u_{-k}
 +\mathcal{O}(u^4).
 \label{z:u_cubic}
\end{align}
This expression can be inverted
\begin{align}
\nonumber
 u_k 
 &={z}_k
 -\frac{1}{4}\sum_{\substack{p_1+p_2=k\\ p_i\neq 0}}\frac{1}{p_1 p_2}{z}_{p_1}{z}_{p_2}
 +\frac{1}{24}\sum_{\substack{p_1+p_2+p_3=k\\p_i\neq 0,k}}\frac{k^2}{ p_1 p_2 p_3(k-p_1)(k-p_2)(k-p_3)}{z}_{p_1}{z}_{p_2}{z}_{p_3}
 \\
 &\quad
 +\frac12\sum_{\substack{p_1+p_2=0\\ p_i\neq 0, k}}\frac{h}{p_1p_2(k-p_1)(k-p_2)[(p_1-p_2)^2+4h]}{z}_{p_1}{z}_{p_2}{z}_k
 -\frac{h(5k^2+h)}{32k^4(k^2+h)^2}{z}_k^2 {z}_{-k}
 +\mathcal{O}({z}^4).
 \label{u:z_cubic}
\end{align}
We are now ready to introduce action and angles variables $I_k, \theta_k$  such that ${24\over c}\omega=\sum_k I_k\, dI_k \wedge d\theta_k$, 
\bea
\frac{{z}_k}{\sqrt{2k(k^2+h)}}=\sqrt{I_k} e^{-i\theta_k}.
\eea
This leads to 
\begin{align}
\label{Ikuk}
    &I_k=\frac{1}{2k (k^2+h)}{z}_k {z}_{-k}
    \\
    &=\frac{1}{2k (k^2+h)} u_k u_{-k}
    +\frac{1}{8k (k^2+h)}\sum_{\substack{p_1+p_2=k\\p_i\neq 0}}\frac{1}{p_1 p_2} u_{p_1} u_{p_2} u_{-k}+ \frac{1}{8k (k^2+h)}\sum_{\substack{p_1+p_2=-k\\p_i\neq 0}}\frac{1}{p_1 p_2} u_{p_1} u_{p_2} u_{k}+\mathcal{O}(u^4). \nonumber
\end{align}

At this point we can go back to $u_0=h-\sum_{n=2}^\infty U_{n}$ are represent it in terms of action variables (by expressing both sides as a series in $z_k$), 
\bea
Q_1\equiv u_0=h+\sum_{k=1}k I_k +\mathcal{O}(z^5).
\eea
This matches the exact relation \eqref{q1} up to the fifth order in $z_k$, reflecting the expansion order in \eqref{u:z_cubic}.

To find $Q_{2n-1}$  in terms of $I_k$ we first write an iterative relation for the Fourier modes of Gelfand-Dikii polynomials, which satisfy \eqref{iterativeR},
\bea
    R_{n,k}&\equiv&\frac{1}{2\pi}\int d \varphi\, e^{-ik\varphi} R_n,\\
  R_{n+1,k}&=&\frac{n+1}{2n+1}\left[2 (k^2+u_0) R_{n,k}
    +Q_{2n-1}u_k
    +\frac{1}{k}\sum_{\ell\neq 0,k}(2k-\ell)u_\ell R_{n,k-\ell}\right] ,
    \quad (k\neq 0). \nonumber
\eea
Then, using the relation between $Q_{2n-1}$ and $R_n$ 
\begin{align}
  R_{n,k} = \frac{1}{ik(2n-1)}\frac{c}{24} \{Q_{2n-1}, u_k\}
\end{align}
we find 
\begin{align}
\nonumber
    \frac{c}{24} \{Q_{2n+1}, u_k\}
    =ik (n+1)Q_{2n-1}u_k+\frac{2(n+1)(k^2+u_0)}{2n-1}\frac{c}{24} \{Q_{2n-1}, u_k\}+ \\ \frac{n+1}{2n-1}\sum_{\ell\neq 0,k}\frac{2k-\ell}{k-\ell}u_\ell\frac{c}{24} \{Q_{2n-1}, u_{k-\ell}\}. \label{Q_rec_rel}
\end{align}
We use the following ansatz for $Q_{2n-1}$ in terms of $u_k$,
\begin{align}
  \label{Q_ansz} 
    Q_{2n-1}=&h^n
    +\frac{1}{2!}\sum_{\substack{p_1+p_2= 0\\p_i\neq0}} q^{(n)}_{p_1,p_2} u_{p_1} u_{p_2}
    +\frac{1}{3!}\sum_{\substack{p_1+p_2+p_3= 0\\p_i\neq0}} q^{(n)}_{p_1,p_2,p_3} u_{p_1} u_{p_2} u_{p_3}
    \\
    &+\frac{1}{4!}\sum_{\substack{p_1+p_2+p_3+p_4= 0\\p_i\neq0}} q^{(n)}_{p_1,p_2,p_3,p_4} u_{p_1} u_{p_2} u_{p_3}u_{p_4}
    +\mathcal{O}(u^5), \nonumber
\end{align}
and 
the iterative relation \eqref{Q_rec_rel} becomes the iterative relation for $q^{(n)}_{p_1,\dots,p_i}$ for $i=2,3,4$,
\begin{align}
\label{qn_2:rr}
    q^{(n+1)}_{k,-k}=\frac{2(n+1)(k^2+h)}{2n-1}q^{(n)}_{k,-k}+\frac{(n+1)h^n}{2(k^2+h)},
\end{align}
\begin{align}
   q^{(n+1)}_{p_1,p_2,p_3}
   &=
   \frac{2(n+1)(p_1^2+p_2^2+p_3^2+3h)}{3(2n-1)}q^{(n)}_{p_1,p_2,p_3}
   -\frac{(n+1)h^n(p_1^2+p_2^2+p_3^2+6h)}{8 (p_1^2+h)(p_2^2+h)(p_3^2+h)}
   \\
   &\quad
   -\frac{n+1}{3(2n-1)}\left[
   \left(\frac{p_1-p_2}{p_2}+\frac{p_1-p_3}{p_3}
   \right) q^{(n)}_{p_1,-p_1} 
   +\text{symmetric w.r.t.  $p_1, p_2 , p_3$}.
   \right]
\end{align}
These can be solved as follows
\begin{align}
    q^{(n)}_{k,-k}=\frac{(h+k^2)^{n-2}(2n)!!}{4(2n-3)!!}\sum_{m=0}^{n-1}\frac{(2m-1)!!}{(2m)!!}\left(\frac{h}{h+k^2}\right)^m,
    \label{qn_2_new}
\end{align}
\begin{align}
\label{q3->q2}
   & q^{(n)}_{p_1,p_2,p_3}=\\
    \
    &\frac{(2n)!!}{8(2n-3)!!p_1 p_2 p_3}\sum_{m=0}^{n-1}\frac{(2m-1)!!}{(2m)!!}h^m\left[p_1(p_1^2+h)^{n-m-2}+p_2(p_2^2+h)^{n-m-2}+p_3(p_3^2+h)^{n-m-2}\right]. \nonumber
\end{align}

Our goal would be to match \eqref{Q_ansz}with the expansion
\begin{align}
    Q_{2n-1}&=h^n +\sum_{k=1}(f^{(n,1)}_k I_k+f^{(n,2)}_{k}I_k^2)+\frac{1}{2}\sum_{\substack{k,\ell=1\\ k \neq \ell}} f^{(n)}_{k,\ell} I_k I_\ell
    +\mathcal{O}(I^3),
\end{align}
by   expressing $I_k$  in terms of $u_k$ using \eqref{Ikuk}. 
This leads to 
\bea
f^{(n,1)}_k=2k (k^2+h) q^{(n)}_{k,-k}
\eea
and the relations for $f^{(n,2)}_k, f^{(n,2)}_{k\ell}$ in terms of $q^{(n)}_{p_1,\dots,p_i}$.  To fix $f^{(2)}_k, f^{(2)}_{k\ell}$, we would not need $q^{(n)}_{p_1,p_2,p_3,p_4}$ with arbitrary $p_1,\dots,p_4$, but only $q^{(n)}_{k,-k,\ell,-\ell}$, including the case of $k=\ell$. The  iterative relation  for $q^{(n)}_{k,-k,\ell,-\ell}$ is cumbersome. Instead, it is more convenient to work directly with the  iterative relation in terms of $f^{(2)}_{k}$ and  $f^{(2)}_{k\ell}$.
Once everything combined together we find
\begin{align}
    f^{(n,1)}_k&=\frac{(2n)!!k(h+k^2)^{n-1}}{2(2n-3)!!}\sum_{m=0}^{n-1}\frac{(2m-1)!!}{(2m)!!}\left(\frac{h}{h+k^2}\right)^m,
    \\
    f_k^{(n,2)}&=
    -\frac{(2n)!!(h+k^2)^{n-2}}{16(2n-3)!!}\sum_{m=0}^{n-1}(3h+k^2 -4 k^2 m)\sum_{j=0}^{m-1}\frac{(2j-1)!!}{(2j)!!}\left(\frac{h}{h+k^2}\right)^j,
\end{align}
and
\bea
&&f^{(n)}_{k,\ell} 
    =\frac{(2n)!! k\ell}{4(2n-3)!!(k^2-\ell^2)}\sum_{j=0}^{n-1}
  (n-1-j)
    \frac{(2j-1)!!h^j}{(2j)!!}\left[
    (h+k^2)^{n-j-1}
    -(h+\ell^2)^{n-j-1}
    \right]
    \\  
    &+&\frac{(2n)!! k\ell}{4(2n-3)!!}
    \sum_{m=0}^{n-1} 
    \sum_{j=0}^{m-1}
    \frac{m(2j-1)!!h^j}{(2j)!!}\left[
    (h+k^2)^{m-j-1}
    (h+\ell^2)^{n-m-1}
    +(h+\ell^2)^{m-j-1}
    (h+k^2)^{n-m-1}
    \right]. \nonumber
\eea
Although written in a different form, this result is in agreement with \eqref{f1}, \eqref{f21}, and \eqref{f22}.

\section{One-zone potentials: details}
\label{A:1cut}
One-zone potentials $u$ can be found from the condition $\{Q_3+\alpha Q_1,u\}=0$ for some constant $\alpha$. From here we immediately find, see section 2.4 of \cite{Dymarsky_2020},
\begin{align}
    \lambda_0&=-\frac{\alpha}{24}-\frac{k^2}{12}(\theta_3(\tau)^4+\theta_4(\tau)^4), \\
    \lambda_1&=-\frac{\alpha}{24}-\frac{k^2}{12}(\theta_2(\tau)^4-\theta_4(\tau)^4), \\
    \lambda_2&=-\frac{\alpha}{24}+ \frac{k^2}{12}(\theta_2(\tau)^4+\theta_3(\tau)^4).
\end{align}
Pertubatively, i.e.~in the limit of small $q=e^{i\pi \tau}$,  corresponding potential is
\begin{align}
    u=h+\frac{32 k^4}{k^2+h}q^2-16 k^2 q \cos(k\varphi)-32 k^2 q^2 \cos(2k\varphi) +\mathcal{O}(q^3).
\end{align}

There are useful relations involving Jacobi elliptic functions and hypergeometric function, 
\bea
\nonumber
&&m:=\theta_2^4(\tau)/\theta_3^4(\tau),\qquad F\left(\frac12,\frac12,1;m\right)=\theta_3(\tau)^2,\quad 
\frac{F\left(\frac12,\frac12,1;1-m\right)}{F\left(\frac12,\frac12,1;m\right)}=-\frac{1}{\pi}\log q,\\
&&\frac{F\left(\frac32,\frac12,1;m\right)}{F\left(\frac12,\frac12,1;m\right)}=1+2 {\partial \ln \theta_3^2(\tau)\over \partial \ln m},\quad 
-16 \sum_{n=0}^\infty\frac{q^{2n+1}}{(1-q^{2n+1})^2}
  +2\theta_3(\tau)^4-2\theta_4(\tau)^4 \frac{F\left(\frac32,\frac12,1;m\right)}{F\left(\frac12,\frac12,1;m\right)}=0. \nonumber
\eea

We also give here more terms in the $q$-expansion of  $I_k$, 
\begin{align}
    I_k=&
    \frac{32 k^3 q^2}{h+k^2}
    +\frac{64 q^4 \left(3 h^2 k^3+12 h k^5+k^7\right)}{\left(h+k^2\right)^3}
    +\frac{128 k^3 q^6 \left(3 h^4+42 h^3 k^2+108 h^2 k^4-58 h
   k^6+k^8\right)}{\left(h+k^2\right)^5} \nonumber
   \\
  & +\frac{128 k^3 q^8 \left(7 h^6+156 h^5 k^2+1083 h^4 k^4+1232 h^3 k^6-4035 h^2 k^8+788 h
   k^{10}+k^{12}\right)}{\left(h+k^2\right)^7}
   +\mathcal{O}(q^{10}), \nonumber
\end{align}
which with help of $Q_1=h+k I_k$ immediately yields
\begin{align}
    Q_1
    &=h+\frac{32 k^4}{k^2+h}q^2+\frac{64k^4(3h^2 +12 h k^2+  k^4) }{(k^2+h)^3}q^4
    \\
    &\quad+\frac{128k^4
    (3h^4+42 h^3 k^2+108 h^2 k^4-58 h k^6+k^8)
    }{(k^2+h)^5}q^6
    +\mathcal{O}(q^7).
\end{align}

The relation for $I_k$ in terms of $q$
can be solved for $q$ in terms of $I_k$ iteratively,  which was used in section \ref{1cut}.
\begin{align}
\nonumber
   q^2=& \frac{ \left(h+k^2\right)}{32 k^3}I_k
   -\frac{\left(3 h^2+12 h k^2+k^4\right)}{512 k^6}I_k^2 
    +\frac{\left(15 h^3+87 h^2 k^2+105 h
   k^4+k^6\right)}{8192 k^9}I_k^3
   \\
  & -\frac{\left(187 h^4+1402 h^3 k^2+3012 h^2 k^4+1606 h k^6+k^8\right)}{262144 k^{12}}I_k^4
   +\mathcal{O}(I_k^5). \nonumber
\end{align}

\section{Perturbative calculation for finite-zone potentials}
\label{perturbative}
We start with the two-zone case and parametrize corresponding differential $dp$ with help of two infinitesimal parameters $\epsilon_1,\epsilon_2$ and $\lambda_0$,
\bea
\label{l1}
\lambda_1&=&\lambda_0+{k^2\over 4}+\epsilon_1+a_1 \epsilon_1^2+ b_1 \epsilon_1 \epsilon_2+ c1 \epsilon_2^2+\dots,\\ 
\lambda_2&=&\lambda_0+{k^2\over 4}-a\epsilon_1+a_2 \epsilon_1^2+ b_2 \epsilon_1 \epsilon_2+ c2 \epsilon_2^2+\dots,\\
\lambda_3&=&\lambda_0+{\ell^2\over 4}+\epsilon_2+a_3 \epsilon_1^2+ b_3 \epsilon_1 \epsilon_2+ c3 \epsilon_2^2+\dots,\\ 
\lambda_4&=&\lambda_0+{\ell^2\over 4}-b\epsilon_2+a_4 \epsilon_1^2+ b_4 \epsilon_1 \epsilon_2+ c4 \epsilon_2^2+\dots,\\
r_1&=&\lambda_0+{k^2\over 4}+d_1 \epsilon_1^2+e_1 \epsilon_1 \epsilon_2 + f_1 \epsilon_2^2+\dots,\\
r_2&=&\lambda_0+{\ell^2\over 4}+d_2 \epsilon_1^2+e_2 \epsilon_1 \epsilon_2 + f_2 \epsilon_2^2+\dots \label{r2}
\eea
The parametrization is redundant, with different choices related by redefinitions of $\epsilon_1,\epsilon_2$. We assume $\epsilon_1\sim \epsilon_2$ are of the same order and in what follows we refer to expansion in $\epsilon_1,\epsilon_2$ simply as $\epsilon$ expansion. While keeping two-zone case in mind for concreteness, most of the discussion below applies to  $m$-zone case with arbitrary $m$. 

\subsection{$a$-cycles}
\label{acycle}
To impose $a_1$-cycle constraint \eqref{zero}, we need to  integrate from $\lambda_1$ to $\lambda_2$. By introducing $x$ via
\bea
\lambda= {\lambda_2+\lambda_1\over 2}+x {\lambda_2-\lambda_1\over 2}, 
\eea
and then expanding in powers of $\epsilon$ we reduce the integral to standard integrals of the form 
\bea
\label{simplei}
\int_{-1}^1 {dx\, x^{2n}\over \sqrt{1-x^2}}= {\sqrt{\pi} \Gamma(n+1/2)\over \Gamma(n+1)}.
\eea
Provided we want to find $Q_{2n-1}$  in terms of $I_k$ by expanding up to $p$-th power, we would need to keep $2p$ terms in $\epsilon$-expansion, up to  and including $\epsilon^{2p}$.
This method works for any $a$-cycle integral and any number of zones. 

\subsection{$b$-cycles}
We start with the $b_1$-cycle, which goes from $\lambda_0$ to $\lambda_1$, and introduce another variable $x$
\bea
\lambda=\lambda_1-x(\lambda_1-\lambda_0).
\eea
We can use the proximity of $\lambda_4$ to $\lambda_3$ to expand $\sqrt{(\lambda-\lambda_3)(\lambda-\lambda_4)}$ in $\epsilon$. 
Now the integral  of interest reduced to a sum of integrals of the form 
\bea
\label{integraloi}
\int_0^1 {dx P(x) \over \sqrt{x(1-x)(16 w+x)}(x-c)^r}
\eea
where $16w$ is a small parameter of order $\epsilon$, 
\bea
16 w={\lambda_2-\lambda_1\over \lambda_1-\lambda_0},
\eea
 $P(x)$ is some polynomial and $c=1-\ell^2/k^2$ (we assumed $\ell>k$). The integral \eqref{integraloi} can be related to 
\bea
\label{defJ}
J_n(c):=\int_0^1 {dx\, x^n \over \sqrt{x(1-x)(16 w +x)}(x-c)}
\eea
by differentiating over $c$.  To evaluate it, it is helpful to first introduce  the integral
\bea
\label{defI}
I_n:=\int_0^1 {dx\, x^n \over \sqrt{x(1-x)(16 w +x)}}=\sum_{m=0}^\infty a_m(n) w^m+ \sum_{m=n}^\infty b_m(n) w^m \ln w,
\eea
which can be expressed as formal series in $w$. Coefficients $a_m(n)$ for $n>m$ and  $b_m(n)$ for any $n,m$ can be found analytically
\bea
\label{amn}
a_m(n)=\frac{(-16)^m \Gamma \left(m+\frac{1}{2}\right) \Gamma (n-m)}{\Gamma (m+1) \Gamma \left(-m+n+\frac{1}{2}\right)},\quad n>m,\\
b_m(n)=\frac{16^m (-1)^{n+1} \Gamma \left(m+\frac{1}{2}\right)}{\Gamma (m+1) \Gamma (m-n+1) \Gamma \left(-m+n+\frac{1}{2}\right)}.
\eea
To find $a_m(n)$ for $m\geq n$ we can use the iterative relation
\bea
I_n=(1-2n)(I_n-I_{n-1})-{1\over 8}\partial_w(I_{n+1}-I_n),
\eea
which follows from the integration by parts, and $a_m(0)$ which can be found directly from \eqref{defI} since the corresponding integral can be evaluated analytically. 
For example we find the following iterative relation for $a_n(n)$,
\bea
a_{m+1}(m+1)=\frac{(-1)^m 2^{2 m+3} \Gamma (2 m+1)}{\Gamma (m+2)^2}-\frac{8 (2 m+1) a_m(m)}{m+1},\quad a_0(0)=0.
\eea 
So far we are interested only in first $2p$ powers of $w$, we only need to worry about $a_m(n)$ with $m\leq 2p$. In our case $p=3$ and we simply 
tabulate values of $a_m(n)$ for $0\leq m\leq 6$ and $m\geq n$ for convenience 
\bea
a_m(n)=\left(
\begin{array}{ccccccc}
 0 & \,\,  & \,\,  & \,\,  & \,\,  & \,\,  & \,\,  \\
 8 & 8 & \,\,  & \,\,  & \,\,  & \,\,  & \,\,  \\
 -84 & -104 & -112 & \,\,  & \,\,  & \,\,  & \,\,  \\
 \frac{2960}{3} & 1152 & \frac{4288}{3} & \frac{4736}{3} & \,\,  & \,\,  & \,\,  \\
 -\frac{37310}{3} & -\frac{42040}{3} & -16368 & -\frac{60992}{3} & -\frac{68224}{3} & \,\,  & \,\,  \\
 \frac{820008}{5} & 180656 & 203584 & \frac{1189248}{5} & \frac{1478144}{5} & \frac{1666048}{5} & \,\,  \\
 -\frac{11153912}{5} & -\frac{12097344}{5} & -\frac{7995904}{3} & -\frac{9011456}{3} & -\frac{17549824}{5} & -\frac{65468416}{15} & -\frac{74166272}{15} \\
\end{array}
\right). \nonumber
\eea

Going back to \eqref{defJ}, we can expand $(x-c)$ in the denominator into power series in $x$, thus reducing the integral to a sum of \eqref{defI}.
Provided $n>2p$ and so far we are only interested in terms of order $w^r$ and $w^r\ln w$ with $r\leq 2p$, only relevant contributions would come from $a_m(n)w^m$ term in \eqref{defI} with $m<n$. Corresponding coefficients are known analytically, \eqref{amn}, and can be re-summed yielding,
\bea
J_n(c)=-\sum_{m=0}^{2p} \frac{(-16)^m \omega ^m \Gamma \left(m+\frac{1}{2}\right) \Gamma (l-m) \, _2F_1\left(1,l-m;l-m+\frac{1}{2};\frac{1}{c}\right)}{c\, \Gamma (m+1)}
+{\mathcal O}(w^{2p+1}). \nonumber
\eea
Here $2F_1$ is {\it regularized} hypergeometric function and this expression is only valid for $n>2p$. To extend it to smaller $n$ we use the 
iterative relation, which follows from the integration by parts,
\bea
J_n={J_{n+1}-I_n\over c}.
\eea
This completes  technical preliminaries as now integral over $b_1$ cycle can be reduced to a number of integrals $J_n$ and their derivatives, so far we are only interested in terms of order $w^r$ with $r\leq 2p$. Clearly, the approach above can be used to evaluate integrals over $b_1$ when there are more than two zones. In this case  
one would need to evaluate integrals 
\bea
\int_0^1 {dx\, x^n \over \sqrt{x(1-x)(16 w +x)}\prod_{i=1}^{m-1}(x-c_i)},
\eea
 where $m$ is the number of zones. This can be reduced to \eqref{defJ} by noting
\bea
\prod_{i=1}^{m-1}{1\over(x-c_i)}=\sum_{i=1}^{m-1} {\alpha_i\over x-c_i},
\eea
with the appropriate coefficients $\alpha_i$. 

To evaluate the integral over $b_2$-cycle from $\lambda_2$ to $\lambda_3$ is more challenging because in the $\epsilon\rightarrow 0$ limit there are singularities appearing 
at both boundaries. There is a straightforward but complicated way. By appropriately changing variables and expanding in $\epsilon$ all terms except for $\sqrt{(\lambda-\lambda_1)(\lambda-\lambda_2)(\lambda-\lambda_3)(\lambda-\lambda_4)}$ we reduce the calculation to  the integral 
\bea
\int_0^1 dx {x^n\over \sqrt{x(1-x)(16 w+x)(1+16 u-x)}}
\eea
for positive small $w,u$. The indefinite integral of this kind can be evaluated analytically. Then the definite integral above can be integrated by expanding it powers of $w,u$ (which both are of order $\epsilon$), and keeping terms up to order $2p$. This is an involved exercise and instead one can use one of the following shortcuts. 

In the particular case of two-zone potential, instead of evaluating integral over $b_2$, one can combine the integral over $b_1$ and $b_2$ such that the contour would enclose $\lambda_0,\dots,\lambda_3$. Now one can deform the contour to go from $\lambda_4$ to infinity, if necessary accompanied by a circle at infinity. At this point integrand can be expanded in $\epsilon$ such that brunch-cut from $\lambda_1$ to $\lambda_2$ disappears, yielding pole singularities at $\lambda=\lambda_0+{k^2/4}$. At this point corresponding integral can be rewritten as 
\bea
\oint_{-\infty}^{-16w} dx {P(x)\over \sqrt{x(1-x)(x+16w)}(x-c)^r},
\eea 
where $P(x)$ is some polynomial and $0\leq c \leq 1$. We also emphasize that to render this integral finite, one may need to close the contour at infinity. 
This integral can be decomposed into a sum of integrals of the form 
\bea
\oint_{-\infty}^{-16w} dx {x^n\over \sqrt{x(1-x)(x+16w)}},
\eea 
and 
\bea
\oint_{-\infty}^{-16w} dx {1\over \sqrt{x(1-x)(x+16w)}(x-c)},
\eea 
and its derivatives. First integral can be reduced to \eqref{defI} by deforming the contour to go from $0$ to $1$. Last integral can be reduced to $J_0$ and $J_1$ with help of modular transformation mapping $\infty$ to $1$, $-16w$ to $0$, and $0$ to $-16 w$.
\bea
x\rightarrow {x+16w\over x-1}.
\eea 

This shortcut works for two-zone case, but with more zones present it is not applicable. Neveftheless there is a very simple trick which make evaluation of $b_2$ and other $b$-cycles unnesessary. Indeed, to satisfy \eqref{zero} and \eqref{integerc} for all cycles, it is sufficient to satisfy \eqref{zero} for all cycles and 
\eqref{integerc} for $b_1$ and also impose that the expansion (\ref{l1}-\ref{r2}), and its generalizations for the case of more than two zones, is invariant under permutation of 
indexes and $k_i$ defined in \eqref{periodicity}. Say, for two zones we find 
\begin{align}
\lambda_1&=\lambda_0+{k^2\over 4}-\epsilon_1+{3\epsilon_1^2\over k^2}+{4\epsilon_2^2 k^2\over \ell^2(k^2-\ell^2)}+{\mathcal O}(\epsilon^3),\\
\lambda_2&=\lambda_0+{k^2\over 4}+\epsilon_1,\\
r_1&=\lambda_0+{k^2\over 4}+{\epsilon_1^2\over 2 k^2}+{2\epsilon_2^2 k^2\over \ell^2(k^2-\ell^2)}+{\mathcal O}(\epsilon^3).
\end{align}
and $\lambda_{3,4},r_2$ related to $\lambda_{1,2},r_1$ by the exchange $\epsilon_1 \leftrightarrow \epsilon_2$ and $k\leftrightarrow \ell$.
The same logic with the permutation symmetry works for any number of zones. 

Above we  only explicitly wrote terms up to $\epsilon^2$, while evaluating all terms up to $\epsilon^6$. 
The simple form of $\lambda_2$ above is a parametrization choice.  
With this choice taking $\epsilon_2=0$ does not close the second zone. One can check that taking 
\bea
\epsilon_2=-\frac{2 \ell^2 \epsilon_1^2}{k^2 \left(k^2-\ell^2\right)}+\dots
\eea
such that $\lambda_4=\lambda_3$ would make $I_\ell$ dicussed below vanish. Alternatively one could chooe $\epsilon_i$ to control the size of $\lambda_{2i}-\lambda_{2i-1}$, 
but with this choice both all $\lambda_i$ would depend on all $\epsilon_i$. 

\subsection{Evaluation of $I_k$, $h$ and $Q_{2n-1}$.}
Evaluation of action variables $I_k$ as a pertubative series in $\epsilon_i$ is straightforward. It is an integral over $a$-cycle and therefore can be evaluated along the lines discussed above. The only difference, in comparision with the discussion in subsection \ref{acycle},  is the term $\ln\lambda$, which needs to be expanded in powers of $\epsilon$ yielding polynomials in $x$ in the numerator of \eqref{simplei},
\bea
I_k={2 \epsilon_1^2\over k(\lambda_0+k^2/4)}+{\mathcal O}(\epsilon^3).
\eea
Again, we only keep terms up to $\epsilon^2$ for simplicity. 

Evaluation of $h$ is also straightforward. To that end one needs to calculate $p(0)$, given by an integral from $0$ to $\lambda_0$. After expanding the integrand in powers of $\epsilon$ it becomes the integral which can be evaluated in a closed form, yielding
\bea
h/4=\lambda_0+\lambda_0\left({2 \epsilon_1^2\over k^2(\lambda_0+k^2/4)}+{2 \epsilon_2^2\over \ell^2(\lambda_0+\ell^2/4)}\right)+{\mathcal O}(\epsilon^3).
\eea

Finally, evaluation of $Q_{2n-1}$ for any given $n$ is also straightforward since $\lambda_i$ are known explicitly. As a result we obtain 
$I_k,h,Q_{2n-1}$ as functions of $\lambda_0$ and $\epsilon_i$. One can then reverse-engineer coefficients in \eqref{genQ} such that it is satisfied.

\section{Spectrum of $Q_{2n-1}$ acting on primaries }
\label{qvac}
In this appendix we outlined calculation of ${\sf Q}^0_{2n}$ \eqref{vacenergy} following \cite{dorey2020geometric}. Starting from the Schr$\ddot{\rm o}$dinger equation \eqref{Schrodinger}, one introduces the following change of variables
\bea
\Psi(x)=E^{l(l-3/2)/4\alpha} w^{(l-3/2)/4} y(w), \qquad x=E^{1\over 2\alpha} w^{1\over 2l}, 
\eea
such that \eqref{Schrodinger} becomes
\bea
-\epsilon^2 \partial_w^2 y+Z(w) y=0,\qquad Z(w)={w},\quad \epsilon=E^{-{\alpha+1\over 2\alpha}}.
\eea
Taking $\epsilon$ as a formal small parameter this equation can be solved via WKB expansion, 
\bea
y(w)=e^{{1\over \epsilon}S(w)},\qquad   -\epsilon S''-S'^2+Z=0,\qquad  S(w)=\sum_{n=0}^\infty \epsilon^n S_n.
\eea
The resulting Riccati equation can be rewritten as the iterative relation to find $S'_n$ with $S_0'=-\sqrt{Z(w)}$. It is more convenient for what follows to make another change of variables $z=w^{\alpha/(l+1/2)}$ and introduce the polynomial ansatz
\bea
S'_n=i{\alpha\over 2l+1}z^{1-{l+1/2\over\alpha}}\tilde{S}_n,\qquad \tilde{S}_n=\sum_{k=0}^n i^n c_k^{(n)} z^{-k+(n-1)(1-1/2\alpha)}(1-z)^{k-(3n-1)/2}
\eea
The Ricatti equation rewritten in terms of $c_k^{(n)}$ gives rise to \eqref{ckn1}, which can be used together with \eqref{ckn0}, to iteratively find $c^{(n)}_k$.  The first few $c_k^{(n)}$ read
\begin{align}
\label{ex-c2}
    c_0^{(2)}=\frac{5}{8}\alpha, \quad 
    c_1^{(2)}=\frac{1}{4}(2\alpha-1), \quad 
    c_2^{(2)}=-\frac{1}{8\alpha}(4 u^2\alpha^2-1).
\end{align}
\begin{align}
c_0^{(3)}=-\frac{15 \alpha ^2}{8},
\quad
c_1^{(3)}=
-\frac{9 \alpha ^2}{4}
+\frac{9 \alpha }{8},
\quad
c_2^{(3)}=\frac{1}{2} \alpha ^2\left(u^2-1\right)
+\frac{3 \alpha }{4}
-\frac{3}{8},
   \quad
   c_3^{(3)}=-\frac{1}{8\alpha}(4 u^2\alpha^2-1)
   .
\end{align}

To obtain ${\sf Q}^0_{2n-1}$ one needs to integrate $\tilde{S}_{2n}(w(z))$  over a Pochhammer contour $\gamma_P$, 
\begin{eqnarray}
{\sf Q}^0_{2n-1}
&=&(-1)^{n} \frac{\Gamma\left(\frac{3}{2}-n-\frac{2 n-1}{2 \alpha}\right)}{\sqrt{\pi} \Gamma\left(1-\frac{2 n-1}{2 \alpha}\right)}
\frac{(2 n-1) \Gamma(n+1)}{4^n(\alpha+1)^{n}} \check{I}_{2 n-1}(\alpha, \hat{l}),\\
 \check{I}_{2n-1}&=&\frac{1}{2(1-e^{\frac{-i\pi (2n-1)}{\alpha}})} \int_{\gamma_P}dz\, \tilde{S}_{2n}(z).
\end{eqnarray}
This integral can be evaluated using, 
\begin{align}
   ( 1-e^{2\pi i a})(1-e^{2\pi i b})B(a,b)= \int_{\gamma_P}dz\, z^{a-1}(1-z)^{b-1},
    \label{Poch-Beta}
\end{align}
where $B(a.b)$ is the Euler beta function. Combining everything together yields \eqref{vacenergy}.

Evaluating ${\sf Q}^0_{2n-1}$ explicitly, using computer algebra to solve for $c_k^{(n)}$ iteratively,  for  small and moderate $n$ is an easy task. To obtain $1/c$ expansion of $\lambda^0_{2n-1}$ for arbitrary $n$ requires knowing corresponding $c_k^{(n)}$ in $1/c$ expansion, i.e.~in the limit of large $\alpha$. This proved to be a difficult task. We obtained first three non-trivial terms of $\lambda^0_{2n-1}$ in  $1/\tilde{c}$ expansion  \eqref{vacuumE}, with the first two terms (\ref{R1},\ref{R2}) in closed analytical form. Functions $y_i$ and $\zeta_i$ there are defined as follows
\begin{eqnarray}
y_1(j)&=&\sum_{\ell=0}^j{1\over 2\ell+1},\\
y_2(j)&=&\sum_{\ell=0}^j{1\over (2\ell+1)^2},\\
\zeta_2(j)&=&\sum_{j_1+j_2=j}\zeta(-2j_1-1)\zeta(-2j_2-1),\\
\zeta_3(j)&=&\sum_{j_1+j_2+j_3=j}\zeta(-2j_1-1)\zeta(-2j_2-1)\zeta(-2j_3-1),
\end{eqnarray}
where sum goes only over non-negative $j_1,j_2,j_3$.  
Third term \eqref{R3} was fixed up to one coefficient $p_j$, with the first several values for $0\leq j \leq 17$ given below 
\begin{eqnarray}
\nonumber
p_j=\left(-\frac{31}{224},\frac{103}{576},-\frac{7883}{21120},\frac{868487}{748800},-\frac{505639}{100800},\frac{394694297}{13708800},-\frac{68117454019}{321753600},\frac{4929720750223}{2540160000},\right. \\ \nonumber \left.  -\frac{199232137825687}{9180864000},  \frac{48745030162337923}{167650560000}, -\frac{618684597383137}{134534400},\frac{7442737871872435019}{87783696000},\right. \\  \nonumber \left.  -\frac{1420749127340184137621}{788237049600},   \frac{46636700018927407368821}{1065512448000},    -\frac{198277953077778046100039}{164670105600}, \right. \\  \nonumber \left. \frac{21869843836862719834306038469}{587058612940800}, -\frac{31428771773709445918185916879}{24404109649920},\right. \\ \left.  \frac{4187283526052269558397574465940213}{84663488093184000},\dots \right). \nonumber
\end{eqnarray}


\bibliographystyle{JHEP}
\bibliography{spectrum}
\end{document}